\documentclass{config}

\begin{document}

\preprint{APS/123-QED}

\title{Reliable confidence regions for quantum tomography using distribution moments}

\author{D.O. Norkin}
\affiliation{Russian Quantum Center, Skolkovo, Moscow 143025, Russia}
\affiliation{National University of Science and Technology ``MISIS”, Moscow 119049, Russia}

\author{E.O. Kiktenko}
\affiliation{Russian Quantum Center, Skolkovo, Moscow 143025, Russia}
\affiliation{National University of Science and Technology ``MISIS”, Moscow 119049, Russia}

\author{A.K. Fedorov}
\affiliation{Russian Quantum Center, Skolkovo, Moscow 143025, Russia}
\affiliation{National University of Science and Technology ``MISIS”, Moscow 119049, Russia}

\date{\today}

\begin{abstract}
    Quantum tomography is a widely applicable method for reconstructing unknown quantum states and processes. However, its applications in quantum technologies usually also require estimating the difference between prepared and target quantum states with reliable confidence intervals.
    In this work, we suggest a computationally efficient and reliable scheme for determining well-justified error bars for quantum tomography.
    We approximate the probability distribution of the Hilbert-Schmidt distance between the target state and the estimation, which is given by the linear inversion, by calculating its two moments. 
    We also present a generalization of this approach for quantum process tomography and deriving confidence intervals for affine functions.
    We benchmark our approach for a number of quantum tomography protocols using both simulation and demonstration with the use of a cloud-accessible quantum processor.
    The obtained results pave the way for the use of the suggested scheme for the complete characterization of quantum systems of various nature.
\end{abstract}

\maketitle

\section{Introduction}

Quantum technologies strongly require the ability to prepare quantum states with desired properties~\cite{Fedorov2022}.
In this regard, one of the crucial tasks is to estimate the difference between the states that are actually prepared and the desired ones, which requires solving the problem of complete characterization of quantum states and processes.
The art of solving this problem is known as quantum tomography, which uses measurement statistics in multiple bases (for a review, see Ref.~\cite{Lvovsky2009}).
Being a gold standard in quantum physics experiments, quantum tomography is routinely performed for systems of various nature, ranging from quantum-optical states~\cite{Lvovsky2001} to trapped-ion quantum simulators~\cite{Blatt2017}. 
However, along with the various benefits, this method encounters a number of challenges. 
The first issue is related to the efficiency of quantum tomography, which becomes unwieldy for large and complex quantum systems, which are required for quantum technologies~\cite{Blume-Kohout2022}. 
A number of methods for improving the performance of quantum tomography protocols have been suggested~\cite{Weinfurter2010,Gross2010,Cramer2010,Blatt2017,Carleo2018,Carrasquilla2019}, 
and the latest results are related to the use of neural-network quantum state representation for tomographic reconstruction~\cite{Carleo2018,Carrasquilla2019,Tiunov2020,Hradil2022,Lvovsky2022}. 

Another important issue is supplying quantum tomography protocols with efficient methods 
for the estimation of their reconstruction accuracy~\cite{Bogdanov2009,Flammia2011,Silva2011,Blume-Kohout2012,Renner2012,Filippov2012,Flammia2012,Sugiyama2013,Renner2016,Renner2019}.  
Most of the existing practical techniques for estimating the accuracy of quantum tomography protocols are based on heuristic approaches,
while only a few methods are established rigorously~\cite{Sugiyama2013,Renner2019,Kiktenko20202,Kiktenko2021}.
Existing approaches for solving this problem encounter an obstacle of combining computational efficiency, which takes place for heuristic methods, and provable accuracy, which is a feature of rigorous methods 
(however, their estimations are typically quite pessimistic so that yielding error bars, which are unnecessarily large). 
Consequently, finding a trade-off between the computational efficiency of the method and provable accuracy is highly important for ongoing experiments. 
This problem becomes even more significant for quantum process tomography (QPT)~\cite{Poyatos1997,Chuang1997,DAriano2001,Kiktenko20202,Kiktenko2021}, 
which is an important tool for characterizing the quality of operations performed by quantum devices~\cite{White2004,Wineland2008}.  

In this work, we present a computationally efficient approach for solving this problem, which demonstrates high performance compared to existing rigorous approaches.
The idea is that the probability distribution of the Hilbert-Schmidt distance between the experimentally prepared state and the estimation, which is given by the linear inversion, can be approximated by calculating its moments.
In particular, the first two moments-based approximation with the gamma function is suggested.
We benchmark our approach for a number of quantum tomography protocols for states and processes using both simulation and data that are generated using a publicly accessible superconducting quantum processor.
As we expect, our results provide a computationally efficient and still justified approach for quantum tomography of states and processes. 

Our work is organized as follows.
In Sec.~\ref{sec:QST}, we provide a short description of the general quantum state tomography protocol (QST) for a finite-dimensional system.
In Sec.~\ref{sec:RQST}, we discuss constructing confidence regions for QST.
In Sec.~\ref{sec:RQPT}, we present a generalization of the method of obtaining confidence regions for QPT.
In Sec.~\ref{sec:affine}, we describe a method of computing confidence intervals in terms of more practical metrics, such as fidelity or mean value of an observable.
In Sec.~\ref{sec:analysis}, we present a performance analysis of our approach, where we use several quantum states, in particular, 3-qubit GHZ state, for the QST, and also single-qubit depolarizing channel, and noisy quantum teleportation channel for the QPT setting.
We conclude in Sec.~\ref{sec:conclusion}.

\section{Quantum state tomography}\label{sec:QST}

Let us first briefly consider a standard QST protocol.
We are interested in a quantum system assigned with a finite-dimensional Hilbert space $\hil$ with a corresponding space $\mathcal{L}(\hil)$ of Hermitian operators over $\hil$.
The quantum system under consideration is repeatedly prepared in the same unknown state, described by a density matrix $\rho$, 
which is an element of the set of density operators $\denset = \left\{\rho\in\mathcal{L}(\hil) \middle\vert\ \rho \ge 0,\ \Tr \rho = 1 \right\} \subset \mathcal{L}(\hil)$. 
To simplify our consideration, we assume that after each preparation the system is measured with informationally complete positive operator-valued measure (POVM), which is defined by a set of Hermitian positive semidefinite operators $\boldsymbol{E}=\povm$ that sum into unity ($\sum_{i=1}^P E_i = \id$) and represent (possibly overcomplete) basis in $\mathcal{L}(\hil)$.
We note that although in practice a set of POVMs is often used in QST protocols instead of a single informationally complete POVM, all further considerations can be straightforwardly generalized to this case as well.
After a series of $N$ such measurements, one obtains data in the form of a vector $\boldsymbol{n}=(n_1,\ldots,n_P)$, where each element $n_i$ indicates the number of occurrences of an outcome corresponding to $E_i$. 
Finally, a density matrix $\tilde\rho$, which is an approximation of the real state $\rho$, can be reconstructed. 
Mathematically, each quantum measurement can be viewed as sampling from a generalized Bernoulli distribution with the probability of $i$th outcome determined as follows: 
\begin{equation}
	p_i = \Tr[E_i\rho].
	\label{eq:prob}
\end{equation} 
Quantum tomography, which is based on a series of such measurements, corresponds to sampling from the multinomial distribution: 
$\boldsymbol{n} \sim \Mult\left(N, \boldsymbol{p}\right)$ with $\boldsymbol{p}=(p_1,\ldots,p_P)$. 
The problem of reconstructing a density matrix thus can be reduced to the task of finding unknown parameters of this distribution.

\section{Constructing confidence regions for the QST protocol}\label{sec:RQST}

To obtain confidence regions for quantum tomography, we first fix the quantum tomography protocol.
Similar to Ref.~\cite{Sugiyama2013}, we use the least squares (LS) inversion that exploits the linearity of Eq.~(\ref{eq:prob}). 
Let us introduce an orthogonal set $\{\sigma_i\}_{i=1}^{d^2-1}$ of traceless Hermitian operators $\sigma_i\in{\cal L}({\cal H})$ satisfying the condition $\Tr(\sigma_i \sigma_j)=d\delta_{ij}$, where $\delta_{ij}$ is a Kronecker delta and $d$ is the dimension of ${\cal H}$.
Each matrix then can be represented by its Pauli vector $\boldsymbol{r}(\rho)=(1/d, r_1(\rho),\ldots, r_{d^2-1}(\rho))$, which satisfies $\rho = \boldsymbol{r}(\rho)\cdot\vec{\sigma}$ with $\vec{\sigma}=(\mathbb{1},\sigma_1,\ldots,\sigma_{d^2-1})$.
One can easily show that there exists a unique matrix $A$ such that 
\begin{equation}\label{eq:state_lin}
	\boldsymbol{p} = A\boldsymbol{r}(\rho).
\end{equation}
It is well known that the LS solution to this equation is $\boldsymbol{r}(\rho) = A^+\boldsymbol{p}$, 
where $A^+ = (A^\top A)^{-1} A^\top$ is a left pseudo-inverse matrix~\cite{Watson1967}. 
Thus, by combining all these formulas, one can derive the expression for $\tilde\rho$ as follows:
\begin{equation}\label{eq:lininv}
	\tilde\rho = \frac{1}{N}\vec{\sigma}\cdot \left(A^+\boldsymbol{n}\right)=
    \vec{\sigma}\cdot \left(A^+\boldsymbol{f}\right),
\end{equation}
where $\boldsymbol{f}:=\boldsymbol{n}/N$ is a vector of observed frequencies, and ${\Tr}\tilde\rho=1$, yet the condition $\tilde\rho\geq 0$ may or may not be fulfilled due to statistical fluctuations.

We are interested in constructing a QST procedure that computes not only a point estimate, but also a confidence region for this estimate with a desired confidence level ${\cal C}$, 
which we denote as $\Gamma(\tilde\rho, {\cal C})$~\cite{Sugiyama2013}. 
This means that $\Pr{\rho\in\Gamma(\tilde\rho, {\cal C})} \ge {\cal C}$, where $\Pr{\cdot}$ should be interpreted as the probability taken over all possible measurement outcomes $\boldsymbol{n}$. 
There are infinitely many regions in various shapes that satisfy the condition above. 
For example, as it is shown, in certain cases the constructed regions have the form of a polytope~\cite{Renner2019, Kiktenko2021}.

In this work, however, we only concentrate on spherically symmetric regions (i.e., having the shape of a ball) with respect to Hilbert-Schmidt norm, centered around the point estimate $\tilde\rho$.
Our task can then be simplified to finding a pair $\delta>0$ and ${\cal C}(\delta)\in(0,1)$ such that the following condition holds:
\begin{equation}
	\Pr{\Delta_{\rm HS}\left(\rho, \tilde\rho\right) < \delta} \ge {\cal C}(\delta),
	\label{eq:pg}
\end{equation}
where the Hilbert-Schmidt distance is defined as
\begin{equation}
	\Delta_{\rm HS}\left(a, b\right) := \sqrt{\frac{\Tr\left[(a - b)^2\right]}{2}}
\end{equation}
for arbitrary Hermitian matrices $a$ and $b$.
Employing the orthogonality of the Pauli basis, one can easily rewrite $\Delta_{\rm HS}\left(\rho, \tilde\rho\right)$ in terms of Pauli vectors:
\begin{equation} \label{eq:hs_norm_est}
    \begin{aligned}
	\Delta_{\rm HS}\left(\tilde\rho, \rho\right) &= \sqrt{\frac{d}{2}} \|\boldsymbol{r}(\tilde\rho)-\boldsymbol{r}(\rho)\|_2 \\&= \sqrt{\frac{d}{2}} \left\|A^+(\frac{\boldsymbol{n}}{N}-\boldsymbol{p})\right\|_2,
    \end{aligned}
\end{equation}
where 
$\|\cdot\|_2$ is a standard $\ell^2$ norm. 
This implies that
\begin{equation}\label{eq:pg_bound}
	\Pr{\Delta_{\rm HS}\left(\rho, \tilde\rho\right) < \delta} = \Pr{\sqrt{\frac{d}{2}}\left\|A^+(\frac{\boldsymbol{n}}{N}-\boldsymbol{p})\right\|_2 < \delta}.
\end{equation}
Thus, the RHS could be used as the lower bound ${\cal C}$ in Eq.~\eqref{eq:pg}. 
Moreover, it is the tightest possible estimate, since Eq.~\eqref{eq:pg} turns into an equality with 
\begin{equation} \label{eq:CL_delta}
    {\cal C}(\delta) = \Pr{\xi_{\boldsymbol{p}}< \frac{2\delta^2}{d} }, \quad
    \xi_{\boldsymbol{p}} := \left\|A^+(\frac{\boldsymbol{n}}{N}-\boldsymbol{p})\right\|_2^2.
\end{equation}

Next, we consider a way of approximating the distribution of $\xi_{\boldsymbol{p}}$.
We note that since we consider a finite-dimensional space both $\|\frac{\boldsymbol{n}}{N}-\boldsymbol{p}\|_2$ and the corresponding operator norm $\left\|A^+\right\|_2$ are bounded. 
This combined with the definition of operator norm implies the boundedness of  $\xi_{\boldsymbol{p}} \leq \left\|A^+\right\|_2^2 \|\frac{\boldsymbol{n}}{N}-\boldsymbol{p}\|_2^2$. 
Thus, its distribution is uniquely determined by its moments (see Ref.~\cite{Kjeldsen1993,Lin2017} on the \emph{moment problem}).

Let us now show that it is feasible to explicitly calculate all moments of $\xi_{\boldsymbol{p}}$.
One can see that
\begin{equation}
    \begin{aligned}
	\xi_{\boldsymbol{p}} &= \sum_i \left(\sum_j\sum_k A^+_{ij}A^+_{ik}\left(\frac{n_j}{N} - p_j\right)\left(\frac{n_k}{N} - p_k\right) \right)\\ 
	&=  \sum_{jk} T_{jk}\left(\frac{n_j}{N} - p_j\right)\left(\frac{n_k}{N} - p_k\right) \\ &= a^{(0)} + \sum_j a^{(1)}_j n_j + \sum_{jk} a^{(2)}_{jk} n_j n_k,
    \end{aligned}
\end{equation}
where  $T:=(A^+)^\top A^+$, $a^{(0)}=\sum_{ij} T_{ij}p_ip_j$, $a^{(1)}_j=\sum_{i} T_{ij}p_i/N$, and $a^{(2)}_{jk}=T_{jk}/N^2$.
As can be seen from the equation above, the integer powers $\xi_{\boldsymbol{p}}^\kappa$ can be expressed as multivariate polynomials of $n_i$, which are samples from a multinomial distribution.
Thus, $\mathbb{E}[\xi_{\boldsymbol{p}}^\kappa]$ could be decomposed into a sum of expected values of monomials containing $n_i$ in degrees up to $2\kappa$ ($\mathbb{E}[n_i]$, $\mathbb{E}[n_in_j]$, $\mathbb{E}[n_i^2n_jn_k]$, and so on). 
This implies that it is feasible to explicitly calculate all moments of $\xi_{\boldsymbol{p}}$, since there are known formulae for said expected values~\cite{newcomer2008computation,stats4010002}.
In particular, the mean and the variance of $\xi_{\boldsymbol{p}}$ are correspondingly given by
\begin{equation}
    \mu_{\boldsymbol{p}} =
    \frac{1}{N}\sum_i T_{ii}p_i,\quad
    V_{\boldsymbol{p}}=
    \frac{1}{N^2}\sum_{i\neq j} T_{ij}^2p_ip_j,
\end{equation}
where only the terms with the highest order in $N^{-1}$ are taken.

In a real experiment, $\boldsymbol{p}$ is unknown, and only a vector of measured frequencies $\tilde{\boldsymbol{f}}$ is available.
Taking into account the fact that the distance $|p_i-\tilde{f}_i|$ scales as $N^{-1/2}$ and that $\mu_{\boldsymbol{p}}$ and $V_{\boldsymbol{p}}$ have a polynomial structure, which bounds their partial derivates with respect to $p_i$, one can take $\mu_{\boldsymbol{p}} \simeq \mu_{\tilde{\boldsymbol{f}}}$ and $V_{\boldsymbol{p}} \simeq 
 V_{\tilde{\boldsymbol{f}}}$ with an introduced relative error of the order of $N^{-1/2}$.
The same considerations can be applied to higher moments or $\xi_{\boldsymbol{p}}$.

Finally, to obtain efficiently computable confidence regions we suggest approximating a distribution of $\xi_{\boldsymbol{p}}$ by gamma distribution with the first two centralized moments given by $\mu_{\tilde{\boldsymbol{f}}}$ and $V_{\tilde{\boldsymbol{f}}}$.
The choice of the gamma distribuition is motivated by its efficency for approximating a weighted sum
of chi-squared random variables~\cite{bodenham2016comparison}.
So we consider an approximation
\begin{equation} \label{eq:gamma_approx}
    {\cal C}(\delta)\approx F_{\mu_{\tilde{\boldsymbol{f}}},V_{\tilde{\boldsymbol{f}}}}(\frac{2\delta^2}{d})
\end{equation}
with the cumulative distribution function (CDF) of the gamma distribution
\begin{equation}
F_{m,\sigma}(x)=
    \frac{
        \gamma\left(m^2/\sigma, xm/\sigma\right)
    }
    { \Gamma(m^2/\sigma)
    },
\end{equation}
where $\gamma(\cdot,\cdot)$ and $\Gamma(\cdot)$ are standard incomplete gamma function and gamma functions correspondingly.
As we show in Sec.~\ref{sec:analysis}, the gamma distribution pretty well mimics the true distribution of $\xi_{\boldsymbol{p}}$.
At the same time, it allows for efficient obtaining of confidence regions without using the Monte-Carlo approach, which is a gold standard in this task~\cite{EfroTibs93,Blume-Kohout2012}.

\section{Constructing confidence regions for QPT}\label{sec:RQPT}

Here we generalize our approach to the task of QPT.
For this purpose consider a completely-positive trace-preserving map (CPTP) also known as a channel $\Phi\!: {\cal L}({\cal H_{\rm in}})\rightarrow {\cal L}({\cal H_{\rm out}})$, 
where ${\cal H_{\rm in}}$ and ${\cal H_{\rm out}}$ are $d_{\rm in}$- and $d_{\rm out}$-dimensional Hilbert spaces ($d_{\rm in}, d_{\rm out}<\infty$), respectively.
In what follows, $\mathbb{1}_{{\rm in}({\rm out})}$ is the identity operator in ${\cal H}_{{\rm in}({\rm out})}$, and $\Tr_{{\rm in}({\rm out})}$ denotes partial trace over ${\cal H}_{{\rm in}({\rm out})}$.

To define the map $\Phi$ it is convenient to consider a corresponding Choi state~\cite{Jamiokowski1972,Choi1975,Jiang2013} 
$C_{\Phi}\in\{C\in {\cal L}({\cal H}_{\rm in}\otimes{\cal H}_{\rm out}): C\geq0, \Tr_{\rm out} C = \mathbb{1}_{\rm in} \}$, given by
\begin{equation}
	C_{\Phi} = \Sum_{i,j=0}^{d_{\rm in}-1} \ket{i}\bra{j}\otimes\Phi\left(\ket{i}\bra{j}\right),
\end{equation}
where $\{\ket{n}\}_{n=0}^{d_{\rm in-1}}$ is a computational basis in ${\cal H}_{\rm in}$.
Recall that given $C_{\Phi}$ one can compute an output of $\Phi$ for any input $\rho\in{\cal L}(\cal H_{\rm in})$ as follows:
\begin{equation}\label{eq:channel_choi}
	\Phi[\rho] = \Tr_{\rm in}( \rho^\top \otimes \mathbb{1}_{\rm out} C_{\Phi}),
\end{equation}
where $\top$ stands for standard transposition.

Consider a QPT protocol, where a set of input states $\boldsymbol\rho_{\rm in}=(\rho_{\rm in}^{(1)},\ldots,\rho_{\rm in}^{(M)})$ with $\rho_{\rm in}^{(i)}\in{\cal S}({\cal H}_{\rm in})$ is repeatedly prepared,
put through $\Phi$, and then is measured by a POVM $\boldsymbol{E}=\povm$ $N$ times.
According to Eq.~\eqref{eq:prob}, the probability of observing $j$-th outcome when $i$-th input state is sent to the channel is $p_{ij} = \Tr\left[\Phi(\rho_{\rm in}^{(i)})E_j\right]$. 
Together with Eq.~\eqref{eq:channel_choi}, it gives us the following formula:
\begin{equation}\label{eq:prob_channel}
	p_{ij} = \Tr\left[E_j\Phi(\rho_{\rm in}^{(i)})\right] = \Tr\left[\left((\rho_{\rm in}^{(i)})^\top\otimes E_j\right)C_{\Phi}\right].
\end{equation}
Let us then denote a Pauli vector of $C_{\Phi}$ by $\boldsymbol{c}(C_{\Phi})$: $C_{\Phi} = \boldsymbol{c}(C_{\Phi}) \cdot \vec{\sigma}_{\rm in,out}$, where
$\vec{\sigma}_{\rm in,out}=({\mathbb{1}\otimes\mathbb{1},\ldots,\sigma^{\rm in}_{d_{\rm in}^2-1}\otimes\sigma^{\rm out}_{d_{\rm out}^2-1}})$ with traceless Hermitian operators satisfying ${\rm Tr}(\sigma_i^{\rm in(out)}\sigma_j^{\rm in(out)})=d_{\rm in(out)}\delta_{ij}$.
Note that $\boldsymbol{c}(C_{\Phi})$ has $d_{\rm in}^2(d_{\rm out}^2-1)$ independent components since ${\rm Tr}_{\rm out}C_{\Phi}=\mathbb{1}_{\rm in}$.

In analogy with Eq.~\eqref{eq:state_lin}, due to linearity there exists a tensor $B_{ijk}$ such that:
\begin{equation}\label{eq:prob_channel_inv}
	\tilde{c}_i = B_{ijk}p_{jk},
\end{equation}
where $\tilde{\boldsymbol{c}}$ is LS inversion estimation of $\boldsymbol{c}(C_{\Phi})$.
Doing analogous calculations one can arrive at very similar expressions to Eq.~\eqref{eq:CL_delta} and Eq.~\eqref{eq:gamma_approx}:
\begin{equation}\label{eq:gamma_approx_channel}
    {\cal C}(\delta)=\Pr{\Delta_{\rm HS}\left(C_{\Phi}, \tilde C_{\Phi}\right) < \delta} \approx F_{\mu_{\tilde{\boldsymbol{f}}},V_{\tilde{\boldsymbol{f}}}}\left(\frac{2\delta^2}{D}\right),
\end{equation}
where the first two moments $\mu_{\tilde{\boldsymbol{f}}}$ and $\mu_{\tilde{\boldsymbol{f}}}$ are calculated as 
\begin{equation}
    \begin{aligned}
    \mu_{\tilde{\boldsymbol{f}}} &= \frac{1}{N}\sum_{jk}\sum_i B_{ijk} B_{ijk} \tilde{f}_{jk},\\
    V_{\tilde{\boldsymbol{f}}} &= \frac{1}{N}\sum_{(j,k)\neq (j',k')}\sum_i B_{ijk} B_{ij'k'} \tilde{f}_{jk} \tilde{f}_{j'k'}
    \end{aligned}
\end{equation}
from the observed frequencies $\tilde{\boldsymbol{f}}$
(here $\tilde C_{\Phi}$ is an estimate of $C_{\Phi}$ obtained via linear inversion and $D=d_{\rm in}d_{\rm out}$).

\section{Deriving confidence intervals for affine functions}\label{sec:affine}

Our method is specifically based on Hilbert-Schmidt norm as a measure of distance between quantum states.
However, often it is desired to compute confidence intervals in terms of more practical metrics, such as fidelity or mean value of an observable.
The method of obtaining confidence intervals for affine functions of quantum states (Choi states) with standard methods of linear programming based on QST (QPT) confidence polytopes is presented in Ref.~\cite{Kiktenko2021}.

Here we demonstrate how to generalize our method to other metrics employing convex optimization.
We consider only the case of QST, as the QPT setting is analogous.
Suppose one has already performed the procedure described in the section devoted to constructing confidence intervals, 
i.e. for a given confidence level ${\cal C}$ calculated $\delta$ such that $\Pr{\Delta_{\rm HS}\left(\rho, \tilde\rho\right) < \delta} \ge {\cal C}$, or in Pauli vector formalism: 
\begin{equation}
	\Pr{\|\boldsymbol{r}(\rho) - \boldsymbol{r}(\tilde\rho)\|_2 < \sqrt{\frac{2}{d}}\delta} \ge {\cal C}.
\end{equation}
Suppose then one wants to compute a confidence interval in terms of an affine function of $\boldsymbol{r}(\rho)$, i.e., a function $\phi: {\cal S}({\cal H})\rightarrow \mathbb{R}$ that can be represented in the following form:
\begin{equation}
	\phi(\rho) = {\boldsymbol r}(\rho)\cdot\boldsymbol\phi + \phi_0,
\end{equation}
where $\boldsymbol\phi\in\mathbb{R}^{d^2}$ and $\phi_0\in\mathbb{R}$.

Thus, the task is to find the optima of $\phi$ in the ball 
\begin{equation}
	\|\boldsymbol{r}(\rho) - \boldsymbol{\tilde r}(\rho)\|_2 < \sqrt{\frac{2}{d}}\delta.
\end{equation} 
This optimization problem can be solved efficiently with second-order cone programming~\footnote{Python Software for Convex Optimization, https://cvxopt.org}. 
The classical second-order cone programming formulation is the following:
\begin{equation}\label{eq:SOCP}
    \begin{aligned}
        &\text{minimize~} {\boldsymbol k}\cdot {\boldsymbol x}\\
        &\text{subject to} \\
        &\qquad \|\boldsymbol{A}_i \boldsymbol{x} + \boldsymbol{b}_i\|_2 \leq \boldsymbol{c}_i\cdot x + d_i, \quad i \in 1,\ldots,m \\
        &\qquad \boldsymbol{F}\boldsymbol{x} = \boldsymbol{g}.
    \end{aligned}
\end{equation}
It is easy to see that our task can be formulated in this classical form with straightforward substitutions.

\section{Performance analysis}\label{sec:analysis}

In this section, we analyze the performance of the developed algorithm in cases of both QST and QPT. 
We consider different quantum states and processes, but utilize the same setup for all of them. 

For probing quantum processes, we use the set of $4^m$ (here and further $m$ denotes the number of qubits in a quantum state or process) input states consisting of pure product states in the form $\rho_{\rm in}^{(i_1,\ldots,i_m)} = \rho_{\rm in}^{(i_1)}\otimes\ldots\otimes\rho_{\rm in}^{(i_m)}$,
where $i_j\in\{1,\ldots,4\}$ and single-qubit states $\{\rho_{\rm in}^{(k)}\}_{k=1}^4$ form a tetrahedron inscribed in a cube with facet orthogonal to $x$, $y$, and $z$ axes.
For read-out in both QST and QPT we use either $3^m$ POVMs corresponding to single-qubit mutually unbiased (MUB) measurements with respect to $x$, $y$, and $z$ axes, or $4^m$-element symmetric, informationally complete POVM (SIC-POVM) with outcomes corresponding to the mentioned tetrahedron employed for states preparation. 
In what follows we refer to these measurements setups as MUB and SIC correspondingly.
In the MUB case, for QST we consider the total number $N$ copies of a quantum state for each of $3^m$ POVMs, and in QPT -- $n$ copies of a quantum process for each of $3^N \times 4^N$ combinations of POVMs and input states.

\begin{figure*}[ht]
    \includegraphics[width=\linewidth]{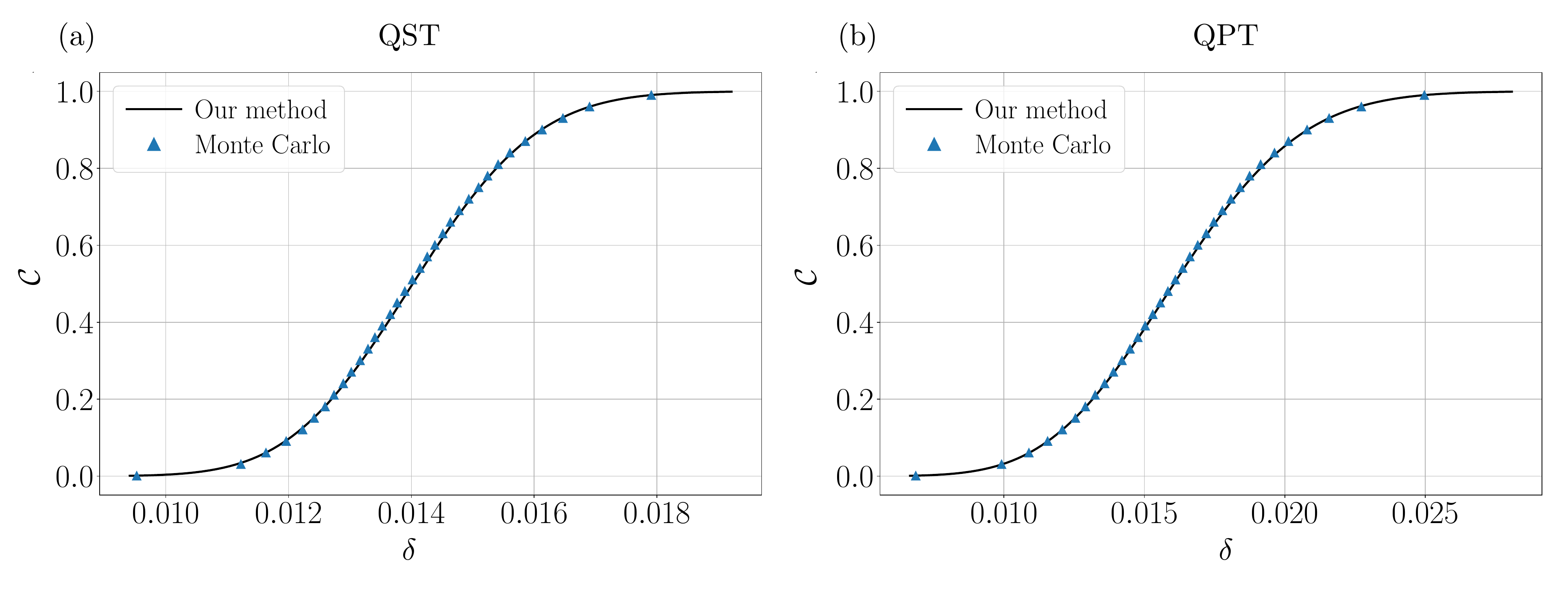}
    \caption{Examples of confidence intervals based on gamma distribution in comparison with Monte-Carlo simulation are presented. 
    In both examples $N$ is set to $10^4$. (a): 3-qubit GHZ state; (b): $\Phi^{(1)}_{0.1}[\rho]$ is the 1-qubit depolarizing channel with $p=0.1$.}
    \label{fig:interval}
\end{figure*}

First, we illustrate how our confidence intervals compare with Monte-Carlo simulations on Fig.~\ref{fig:interval} in both QST and QPT settings. 
In case of QST, we consider a 3-qubit GHZ state $\ket{\psi^{(3)}} = 2^{-1/2}(\ket{000} + \ket{111})$.
For QPT we use a single-qubit depolarizing channel $\Phi^{(m)}_p[\rho] = (1-p)\rho + \frac{\mathbb{1}}{2^m}\Tr \rho$, where $m=1$ and $p=0.1$. 
We note that our intervals almost exactly coincide with those obtained by Monte-Carlo simulation, despite the latter being significantly more computationally intensive.

Then we provide a validation of our confidence intervals. 
We pick a set of quantum states and quantum processes, a set of predefined confidence levels, and for each combination of them do the following procedure.
We perform a simulated quantum tomography with MUB measurements, reconstruct the quantum state (or process) from the results of the tomography, and calculate the confidence interval with respect to the corresponding confidence level. 
After that, we check whether the reconstructed quantum state (or process) lies inside the confidence interval. 
Then we repeat this procedure $R=10^4$ times. 
In the ideal case, if the confidence interval with a confidence level ${\cal C}$ is valid and the most tight possible, the fraction of measurements $f_{\rm in}({\cal C})$, in which the reconstructed state (or process) lies inside the interval should be exactly equal to ${\cal C}$ (in the limit $R\rightarrow\infty)$. 
The numerical results of our verification for QST and QPT can be found in Tables~\ref{tab:verification_qst} and \ref{tab:verification_qpt}, respectively. 
As it is seen, regardless of the quantum state or process and the confidence level our confidence intervals are very close to ideal.

\begin{table}[ht]
    \centering
    \begin{tabular}{c|ccccc}
        State & 0.5 & 0.75 & 0.9 & 0.95 & 0.99 \\ \hline
        $\ket{0}$ & 0.4985 & 0.7532 & 0.9006 & 0.9509 & 0.9899 \\
        $\cos{\frac{\pi}{8}}\ket{0} + \sin{\frac{\pi}{8}} e^{\frac{\imath\pi}{4}} \ket{1}$ & 0.4916 & 0.7503 & 0.9010 & 0.9521 & 0.9911 \\
        Fully mixed 1-qubit & 0.5002 & 0.7532 & 0.9014 & 0.9526 & 0.9896 \\
        $\ket{00}$ & 0.5043 & 0.7582 & 0.9034 & 0.9506 & 0.9885 \\
        $2^{-1/2}(\ket{00}+\ket{11})$ & 0.5065 & 0.7514 & 0.8968 & 0.9436 & 0.9873 \\
        Fully mixed 2-qubit & 0.4977 & 0.7481 & 0.8987 & 0.9483 & 0.9878 \\
        3-qubit GHZ & 0.4967 & 0.7510 & 0.8990 & 0.9501 & 0.9895 \\
    \end{tabular}
    \caption{Verification of confidence intervals for QST.}
    \label{tab:verification_qst}
\end{table}

\begin{table}[ht]
    \centering
    \begin{tabular}{c|ccccc}
        Process & 0.5 & 0.75 & 0.9 & 0.95 & 0.99 \\ \hline
        Hadamard gate & 0.5028 & 0.7656 & 0.9018 & 0.9474 & 0.9867 \\
        $RX(\frac{\pi}{2})$ & 0.5096 & 0.7575 & 0.9021 & 0.9494 & 0.9872 \\
        $RY(\frac{\pi}{2})$ & 0.5133 & 0.7691 & 0.9088 & 0.9547 & 0.9893 \\
        $\Phi^{(1)}_{0.1}[\rho]$ & 0.5070 & 0.7566 & 0.9000 & 0.9485 & 0.9869 \\
        $\Phi^{(2)}_{0.1}[\rho]$ & 0.5162 & 0.7590 & 0.9017 & 0.9493 & 0.9864 \\
    \end{tabular}
    \caption{Verification of confidence intervals for QPT.}
    \label{tab:verification_qpt}
\end{table}

\begin{figure*}
    \centering  \includegraphics[width=\linewidth]{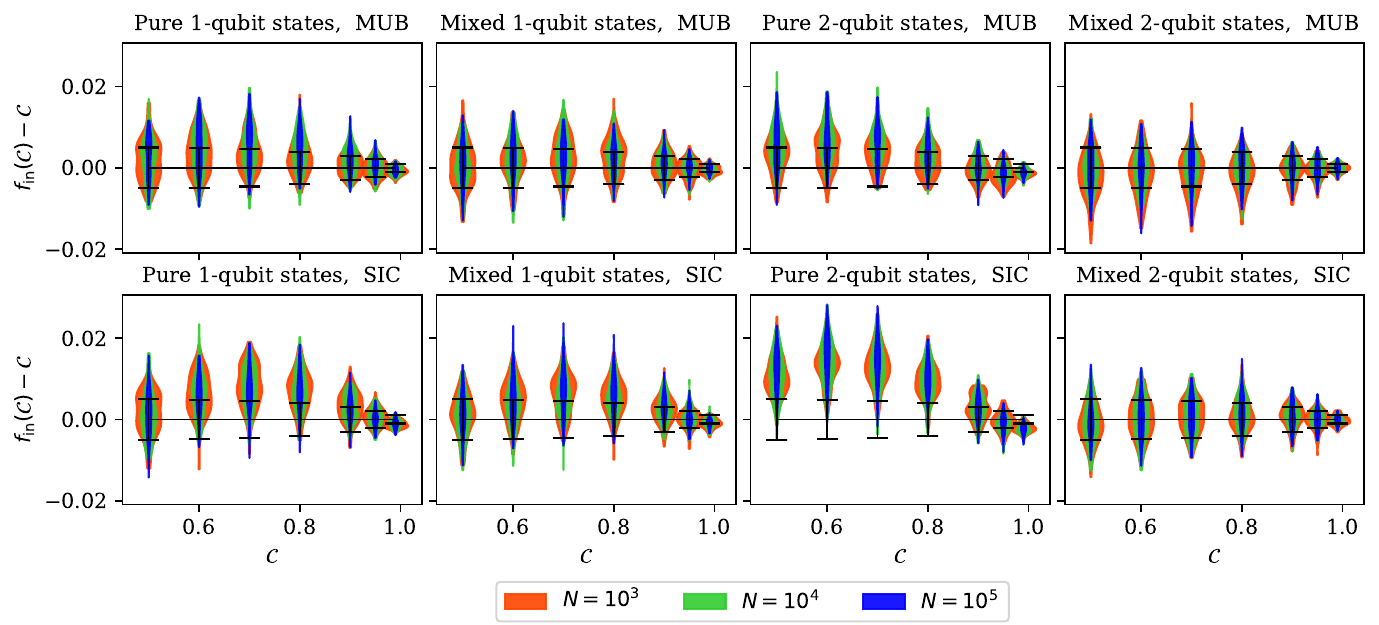}
    \caption{
    Distributions of $f_{\rm in}({\cal C})-{\cal C}$ for various configurations of QST experiments with respect to random pure and mixed states single- and two-qubit states.
    Error bars represent an expected effective spread $\pm\sqrt{{\cal C}(1-{\cal C})/R}$, which appears due to the finite number of samples $R$ used for calculating $f_{\rm in}({\cal C})$.
    }
    \label{fig:violin_qst}
\end{figure*}

\begin{figure*}
    \centering  \includegraphics[width=\linewidth]{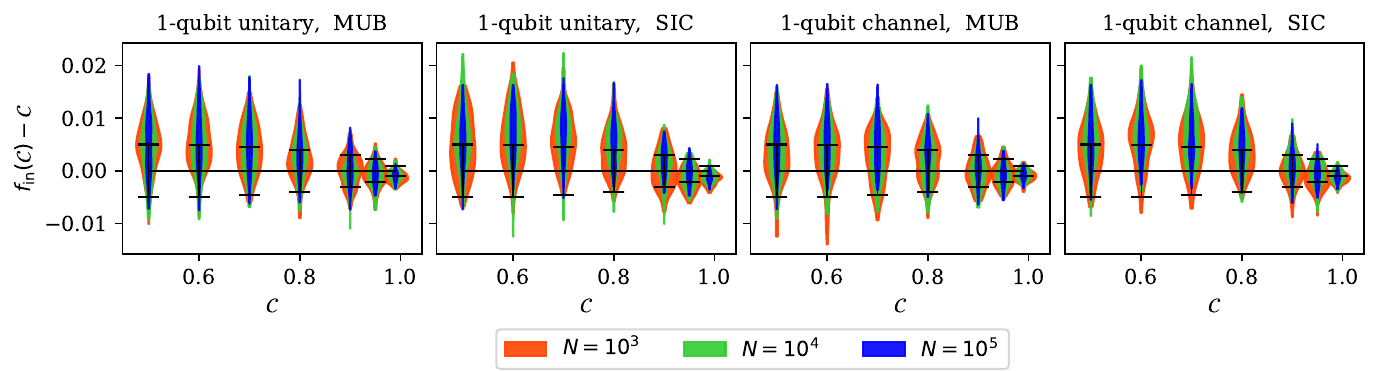}
    \caption{
    Distributions of $f_{\rm in}({\cal C})-{\cal C}$ for various configurations of QPT experiments with respect to random single-qubit unitaries and channels.
    As in Fig.~\ref{fig:violin_qst}, error bars represent an expected effective spread $\pm\sqrt{{\cal C}(1-{\cal C})/R}$, which appears due to the finite number of samples $R$ used for calculating $f_{\rm in}({\cal C})$.
    }
    \label{fig:violin_qpt}
\end{figure*}

To provide an additional validation, we perform a number of numerical experiments with respect to random single-qubit and two-qubit states, and random single-qubit processes.
In particular, we consider Haar-random pure states, Hilbert-Schmidt uniform mixed states~\cite{zyczkowski2001induced}, Haar random unitary operators, and single-qubit quantum channels obtained by using Haar-random two-qubit unitaries and taking a partial trace over an ancillary qubit.
We also vary the number of measurements $N=10^3,10^4,10^5$.
For each configuration $L=100$ random states or processes are generated, and for each of them $R=10^4$ tomography experiments are performed in order to obtain $f_{\rm in}({\cal C})$.
The resulting distribution of differences $f_{\rm in}({\cal C})-{\cal C}$ for QST and QPT are shown in Fig.~\ref{fig:violin_qst} and Fig.~\ref{fig:violin_qpt} respectively.
One can see that magnitudes of observed differences between $f_{\rm in}({\cal C})$ and ${\cal C}$ do not exceed several percents, that makes the suggested method to be suitable for practical purposes.
We also note the case of $f_{\rm in}({\cal C})-{\cal C}>0$ satisfies the required inequality~\eqref{eq:pg} for the confidence level.

To highlight the computational efficiency of our method we report the results of profiling it against the Monte-Carlo approach with 1000 bootstrap samples.
In Table~\ref{tab:profiling} we show time it took to construct a confidence interval depending on the chosen method both for QST with $\ket{\psi^{(3)}}$ and QPT with $\Phi^{(2)}_{0.1}[\rho]$. 
All simulations were conducted on the same machine with a Hexa-Core AMD Ryzen 5 and 16 GB of RAM.

\begin{table}[ht]
    \centering
    \begin{tabular}{c|cc}
         & QST ($\ket{\psi^{(3)}}$) & QPT ($\Phi^{(2)}_{0.1}[\rho]$) \\ \hline
        Our method & 220~ms & 640~ms \\
        Monte-Carlo & 8.5~s & 870~s \\
    \end{tabular}
    \caption{Profiling of confidence regions.}
    \label{tab:profiling}
\end{table}

\begin{figure}[ht]
    \includegraphics[width=.99\linewidth]{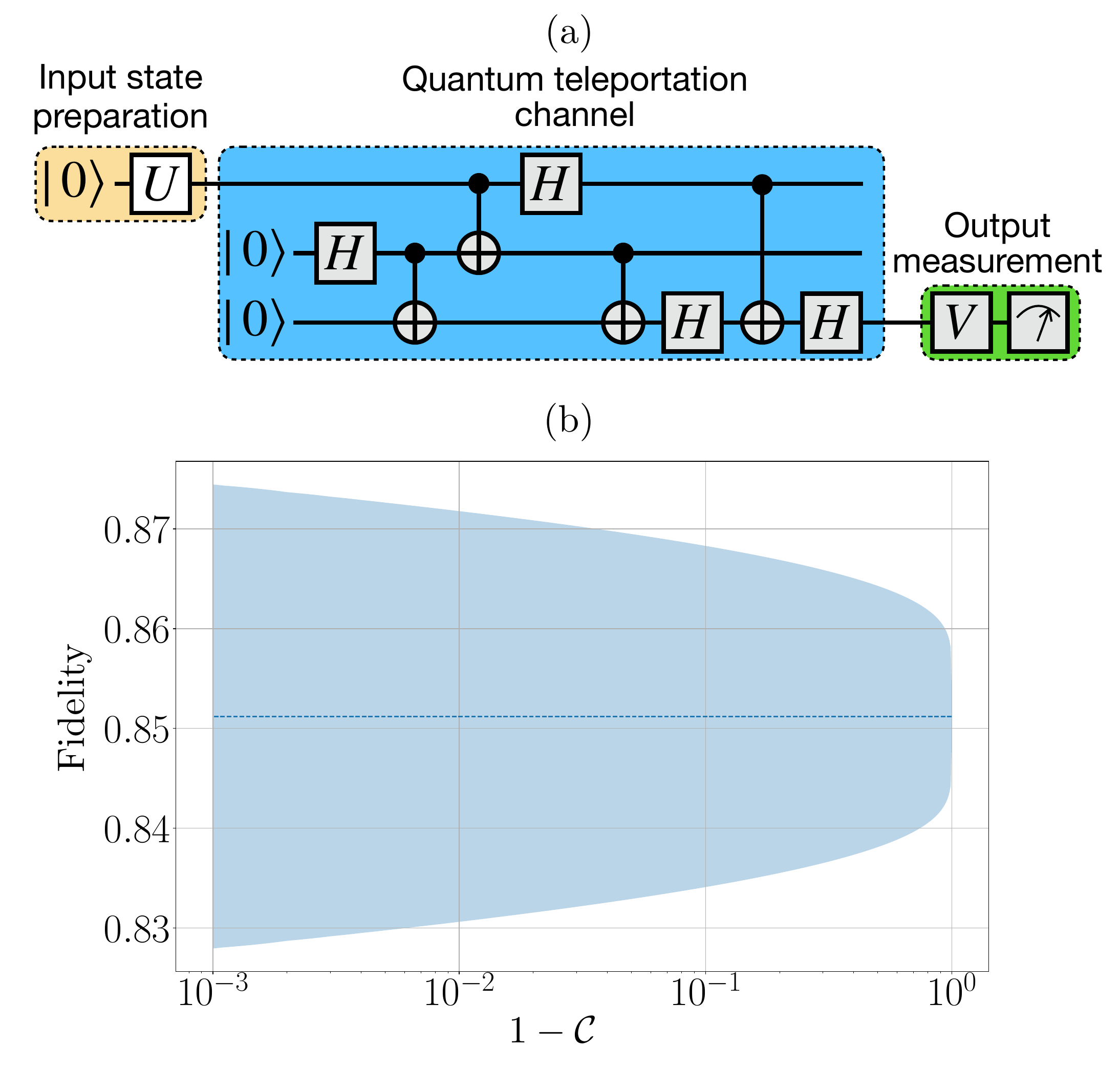}
    \caption{
        In (a) the circuit of QPT protocol of a quantum teleportation channel used in the demonstration is depicted.
        Here standard notations for Hadamard gate, Controlled-NOT gate, and computational basis read-out measurement are used (fully connected qubits ${\sf Q}_0$, ${\sf Q}_1$, ${\sf Q}_2$ of ${\sf ibmqx2}$ backend were taken).
        In (b) the resulting fidelity confidence intervals for QPT  from run on IBM quantum experience processor is demonstrated.
        }
    \label{fig:fidelity}
\end{figure}

We also provide an example of obtaining the fidelity confidence intervals. 
For this purpose, we process the data obtained from the superconducting quantum processor ${\sf ibmqx2}$ (${\sf IBMQ~5~Yorktown}$) provided by IBM (for more details see Appendix~\ref{app:calibration} and~\cite{IBMQ5}).
The circuit used to build the channel is shown on Fig.~\ref{fig:fidelity}a and corresponds to the quantum teleportation of the 1st qubit onto the 3rd qubit.
To prepare input states, an operation $U$ was consequently taken in the form
\begin{equation}
    U = \mathbb{1},~RX(\theta),~ RZ(2\phi)RX(\theta),~RZ(\phi)RY(\theta),
\end{equation}
where $RX$, $RY$, and $RZ$ stand for standard Pauli rotations, $\phi:=2\pi/3$, $\theta:=\arccos(-1/3)$. 
Informationally complete set of output measurements was realized by taking
\begin{equation}
    V=\mathbb{1},~RX(\pi/2),~RY(\pi/2).
\end{equation}

In case of ideal input states, gates and output measurements, this channel should coincide with the identity channel. 
However, due to noise and imperfections real results differ from theory.
On Fig.~\ref{fig:fidelity}b a confidence interval in terms of fidelity with respect to the 1-qubit identity channel for the case of $N=2^{13}\approx 8\times10^3$ measurements for each configuration of an input state and a POVM.
We see that the results of the methods can be used as lower and upper bounds on the performance of NISQ devices.

\section{Conclusion}\label{sec:conclusion}

We have proposed and tested the computationally efficient and reliable scheme for determining well-justified error bars for quantum tomography.
We have restricted ourselves to the use of two first moments of the distribution of $\|\xi_{\boldsymbol{p}}\|_2^2$ only. 
However, as shown above, it is possible to calculate an arbitrarily high moment. 
This opens a broad field of investigation of more precise approximations. 
One may use, for example, the generalized gamma distribution, estimating its parameters with three first moments. 
Overall, the more moments are used, the more accurate the estimation of intervals is expected to be.
A way to develop this technique further is to expand the range of quantum tomography protocols, 
including coherent-state quantum tomography~\cite{Lvovsky2008,Lvovsky2011,Anis2012,Fedorov2015, ghalaii2017scheme} and tomography of state present in setups with postselection~\cite{kiktenko2023exploring}.

{\bf Acknowledgements.}
We acknowledge use of the IBM Q Experience for this work. 
The views expressed are those of the authors and do not reflect the official policy or position of IBM or the IBM Q Experience team. 
The demonstration based on the IBM Q Experience that is used in this paper has been completed in 2021.
We thank M.D. Sapova and A.A. Karazeev for
fruitful discussions.
The research is supported by the Priority 2030 program at the National University of Science and Technology ``MISIS'' under the project K1-2022-027.
The work of E.O.K. (deriving confidence regions) was also supported by the RSF Grant No. 19-71-10091.

\appendix

\section{Cloud platform details} \label{app:calibration}

\begin{figure}
    \centering
    \includegraphics[width=0.45\linewidth]{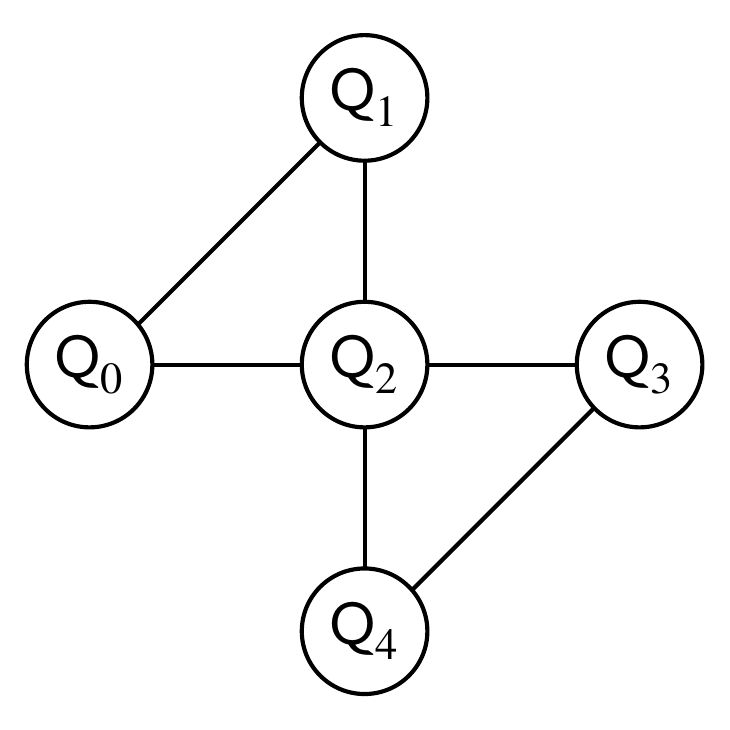}
    \caption{Coupling map of ${\sf ibmqx2}$ ({\sf IBM Q 5 Yorktown}) processor. 
    Connections between qubits ${\sf Q}_i$ stand for a possibility to apply a controlled-NOT gate.}
    \label{fig:coupling}
\end{figure}

In Table~\ref{tab:calibration_data} we provide the calibration data of ${\sf ibmqx2}$ ({\sf IBM Q 5 Yorktown}) processor employed for demonstration in Sec.~\ref{sec:analysis}.
The coupling map of the processor is shown in Fig.~\ref{fig:coupling}
All the data is retrieved from version 1.1.0 information in~\cite{IBMQ5} (the demonstration was performed in May 2018). 
Here, $\omega_i^{\rm R}$ is the resonance frequency of the readout resonator, $\omega_i$ is the frequency of the $i$-th qubit, $\delta_i$ is the anharmonicity (the difference between the frequency of the $\ket{1}$-$\ket{2}$ and the $\ket{0}$-$\ket{1}$ transition), 
$\chi$ is the qubit-cavity coupling strength, $\kappa$ is the cavity coupling to the environment, and $\xi_{ij}$ is the crosstalk matrix.

\begin{table*}[]
    \centering    
    \begin{tabular}{||c|c|c|c|c|c||}
        Qubit & $\omega_i^{\rm R}/2\pi$ (GHz) & $\omega_i/2\pi$ (GHz) & $\delta_i/2\pi$ (MHz) & $\chi/2\pi$ (KHz) & $\kappa/2\pi$ (KHz) \\ \hline
${\sf Q}_0$ & 6.53051 & 5.2760 & -330.3 & 476 & 523 \\
${\sf Q}_1$ & 6.48165 & 5.2122 & -331.9 & 395 & 489 \\
${\sf Q}_2$ & 6.43617 & 5.0154 & -331.2 & 428 & 415 \\
${\sf Q}_3$ & 6.57952 & 5.2805 & -329.4 & 412 & 515 \\
${\sf Q}_4$ & 6.53023 & 5.0711 & -335.5 & 339 & 480 
    \end{tabular}\quad
    \begin{tabular}{||c|c|c|c|c|c||}
        $\xi_{ij}/2\pi$ (kHz) & ${\sf Q}_0$ & ${\sf Q}_1$ & ${\sf Q}_2$ & ${\sf Q}_3$ & ${\sf Q}_4$ \\\hline
        ${\sf Q}_0$ &  & -43 & -83 & &\\
        ${\sf Q}_1$ & -45 &  & -25 & &\\
        ${\sf Q}_2$ & -83 & -27 &  & -127 &  -38\\
        ${\sf Q}_3$ &  & & -127 &  & -97\\
        ${\sf Q}_4$ &  & &-34 &-97 & \\ 
    \end{tabular}
    
    \caption{Calibration data for the five-qubit superconducting processor ${\sf ibmqx2}$ ({\sf IBM Q 5 Yorktown}) at the time of the demonstration (see version 1.1.0 in~\cite{IBMQ5}).}
    \label{tab:calibration_data}
\end{table*}

\bibliography{bibliography.bib}

\begin{thebibliography}{48}%
\makeatletter
\providecommand \@ifxundefined [1]{%
 \@ifx{#1\undefined}
}%
\providecommand \@ifnum [1]{%
 \ifnum #1\expandafter \@firstoftwo
 \else \expandafter \@secondoftwo
 \fi
}%
\providecommand \@ifx [1]{%
 \ifx #1\expandafter \@firstoftwo
 \else \expandafter \@secondoftwo
 \fi
}%
\providecommand \natexlab [1]{#1}%
\providecommand \enquote  [1]{``#1''}%
\providecommand \bibnamefont  [1]{#1}%
\providecommand \bibfnamefont [1]{#1}%
\providecommand \citenamefont [1]{#1}%
\providecommand \href@noop [0]{\@secondoftwo}%
\providecommand \href [0]{\begingroup \@sanitize@url \@href}%
\providecommand \@href[1]{\@@startlink{#1}\@@href}%
\providecommand \@@href[1]{\endgroup#1\@@endlink}%
\providecommand \@sanitize@url [0]{\catcode `\\12\catcode `\$12\catcode
  `\&12\catcode `\#12\catcode `\^12\catcode `\_12\catcode `\%12\relax}%
\providecommand \@@startlink[1]{}%
\providecommand \@@endlink[0]{}%
\providecommand \url  [0]{\begingroup\@sanitize@url \@url }%
\providecommand \@url [1]{\endgroup\@href {#1}{\urlprefix }}%
\providecommand \urlprefix  [0]{URL }%
\providecommand \Eprint [0]{\href }%
\providecommand \doibase [0]{http://dx.doi.org/}%
\providecommand \selectlanguage [0]{\@gobble}%
\providecommand \bibinfo  [0]{\@secondoftwo}%
\providecommand \bibfield  [0]{\@secondoftwo}%
\providecommand \translation [1]{[#1]}%
\providecommand \BibitemOpen [0]{}%
\providecommand \bibitemStop [0]{}%
\providecommand \bibitemNoStop [0]{.\EOS\space}%
\providecommand \EOS [0]{\spacefactor3000\relax}%
\providecommand \BibitemShut  [1]{\csname bibitem#1\endcsname}%
\let\auto@bib@innerbib\@empty
\bibitem [{\citenamefont {Fedorov}\ \emph {et~al.}(2022)\citenamefont
  {Fedorov}, \citenamefont {Gisin}, \citenamefont {Beloussov},\ and\
  \citenamefont {Lvovsky}}]{Fedorov2022}%
  \BibitemOpen
  \bibfield  {author} {\bibinfo {author} {\bibfnamefont {A.~K.}\ \bibnamefont
  {Fedorov}}, \bibinfo {author} {\bibfnamefont {N.}~\bibnamefont {Gisin}},
  \bibinfo {author} {\bibfnamefont {S.~M.}\ \bibnamefont {Beloussov}}, \ and\
  \bibinfo {author} {\bibfnamefont {A.~I.}\ \bibnamefont {Lvovsky}},\
  }\href@noop {} {\enquote {\bibinfo {title} {Quantum computing at the quantum
  advantage threshold: a down-to-business review},}\ } (\bibinfo {year}
  {2022}),\ \Eprint {http://arxiv.org/abs/2203.17181} {arXiv:2203.17181
  [quant-ph]} \BibitemShut {NoStop}%
\bibitem [{\citenamefont {Lvovsky}\ and\ \citenamefont
  {Raymer}(2009)}]{Lvovsky2009}%
  \BibitemOpen
  \bibfield  {author} {\bibinfo {author} {\bibfnamefont {A.~I.}\ \bibnamefont
  {Lvovsky}}\ and\ \bibinfo {author} {\bibfnamefont {M.~G.}\ \bibnamefont
  {Raymer}},\ }\href {\doibase 10.1103/RevModPhys.81.299} {\bibfield  {journal}
  {\bibinfo  {journal} {Rev. Mod. Phys.}\ }\textbf {\bibinfo {volume} {81}},\
  \bibinfo {pages} {299} (\bibinfo {year} {2009})}\BibitemShut {NoStop}%
\bibitem [{\citenamefont {Lvovsky}\ \emph {et~al.}(2001)\citenamefont
  {Lvovsky}, \citenamefont {Hansen}, \citenamefont {Aichele}, \citenamefont
  {Benson}, \citenamefont {Mlynek},\ and\ \citenamefont
  {Schiller}}]{Lvovsky2001}%
  \BibitemOpen
  \bibfield  {author} {\bibinfo {author} {\bibfnamefont {A.~I.}\ \bibnamefont
  {Lvovsky}}, \bibinfo {author} {\bibfnamefont {H.}~\bibnamefont {Hansen}},
  \bibinfo {author} {\bibfnamefont {T.}~\bibnamefont {Aichele}}, \bibinfo
  {author} {\bibfnamefont {O.}~\bibnamefont {Benson}}, \bibinfo {author}
  {\bibfnamefont {J.}~\bibnamefont {Mlynek}}, \ and\ \bibinfo {author}
  {\bibfnamefont {S.}~\bibnamefont {Schiller}},\ }\href {\doibase
  10.1103/PhysRevLett.87.050402} {\bibfield  {journal} {\bibinfo  {journal}
  {Phys. Rev. Lett.}\ }\textbf {\bibinfo {volume} {87}},\ \bibinfo {pages}
  {050402} (\bibinfo {year} {2001})}\BibitemShut {NoStop}%
\bibitem [{\citenamefont {Lanyon}\ \emph {et~al.}(2017)\citenamefont {Lanyon},
  \citenamefont {Maier}, \citenamefont {Holz{\"a}pfel}, \citenamefont
  {Baumgratz}, \citenamefont {Hempel}, \citenamefont {Jurcevic}, \citenamefont
  {Dhand}, \citenamefont {Buyskikh}, \citenamefont {Daley}, \citenamefont
  {Cramer}, \citenamefont {Plenio}, \citenamefont {Blatt},\ and\ \citenamefont
  {Roos}}]{Blatt2017}%
  \BibitemOpen
  \bibfield  {author} {\bibinfo {author} {\bibfnamefont {B.~P.}\ \bibnamefont
  {Lanyon}}, \bibinfo {author} {\bibfnamefont {C.}~\bibnamefont {Maier}},
  \bibinfo {author} {\bibfnamefont {M.}~\bibnamefont {Holz{\"a}pfel}}, \bibinfo
  {author} {\bibfnamefont {T.}~\bibnamefont {Baumgratz}}, \bibinfo {author}
  {\bibfnamefont {C.}~\bibnamefont {Hempel}}, \bibinfo {author} {\bibfnamefont
  {P.}~\bibnamefont {Jurcevic}}, \bibinfo {author} {\bibfnamefont
  {I.}~\bibnamefont {Dhand}}, \bibinfo {author} {\bibfnamefont {A.~S.}\
  \bibnamefont {Buyskikh}}, \bibinfo {author} {\bibfnamefont {A.~J.}\
  \bibnamefont {Daley}}, \bibinfo {author} {\bibfnamefont {M.}~\bibnamefont
  {Cramer}}, \bibinfo {author} {\bibfnamefont {M.~B.}\ \bibnamefont {Plenio}},
  \bibinfo {author} {\bibfnamefont {R.}~\bibnamefont {Blatt}}, \ and\ \bibinfo
  {author} {\bibfnamefont {C.~F.}\ \bibnamefont {Roos}},\ }\href {\doibase
  10.1038/nphys4244} {\bibfield  {journal} {\bibinfo  {journal} {Nature
  Physics}\ }\textbf {\bibinfo {volume} {13}},\ \bibinfo {pages} {1158}
  (\bibinfo {year} {2017})}\BibitemShut {NoStop}%
\bibitem [{\citenamefont {Proctor}\ \emph {et~al.}(2022)\citenamefont
  {Proctor}, \citenamefont {Rudinger}, \citenamefont {Young}, \citenamefont
  {Nielsen},\ and\ \citenamefont {Blume-Kohout}}]{Blume-Kohout2022}%
  \BibitemOpen
  \bibfield  {author} {\bibinfo {author} {\bibfnamefont {T.}~\bibnamefont
  {Proctor}}, \bibinfo {author} {\bibfnamefont {K.}~\bibnamefont {Rudinger}},
  \bibinfo {author} {\bibfnamefont {K.}~\bibnamefont {Young}}, \bibinfo
  {author} {\bibfnamefont {E.}~\bibnamefont {Nielsen}}, \ and\ \bibinfo
  {author} {\bibfnamefont {R.}~\bibnamefont {Blume-Kohout}},\ }\href {\doibase
  10.1038/s41567-021-01409-7} {\bibfield  {journal} {\bibinfo  {journal}
  {Nature Physics}\ }\textbf {\bibinfo {volume} {18}},\ \bibinfo {pages} {75}
  (\bibinfo {year} {2022})}\BibitemShut {NoStop}%
\bibitem [{\citenamefont {T\'oth}\ \emph {et~al.}(2010)\citenamefont {T\'oth},
  \citenamefont {Wieczorek}, \citenamefont {Gross}, \citenamefont {Krischek},
  \citenamefont {Schwemmer},\ and\ \citenamefont
  {Weinfurter}}]{Weinfurter2010}%
  \BibitemOpen
  \bibfield  {author} {\bibinfo {author} {\bibfnamefont {G.}~\bibnamefont
  {T\'oth}}, \bibinfo {author} {\bibfnamefont {W.}~\bibnamefont {Wieczorek}},
  \bibinfo {author} {\bibfnamefont {D.}~\bibnamefont {Gross}}, \bibinfo
  {author} {\bibfnamefont {R.}~\bibnamefont {Krischek}}, \bibinfo {author}
  {\bibfnamefont {C.}~\bibnamefont {Schwemmer}}, \ and\ \bibinfo {author}
  {\bibfnamefont {H.}~\bibnamefont {Weinfurter}},\ }\href {\doibase
  10.1103/PhysRevLett.105.250403} {\bibfield  {journal} {\bibinfo  {journal}
  {Phys. Rev. Lett.}\ }\textbf {\bibinfo {volume} {105}},\ \bibinfo {pages}
  {250403} (\bibinfo {year} {2010})}\BibitemShut {NoStop}%
\bibitem [{\citenamefont {Gross}\ \emph {et~al.}(2010)\citenamefont {Gross},
  \citenamefont {Liu}, \citenamefont {Flammia}, \citenamefont {Becker},\ and\
  \citenamefont {Eisert}}]{Gross2010}%
  \BibitemOpen
  \bibfield  {author} {\bibinfo {author} {\bibfnamefont {D.}~\bibnamefont
  {Gross}}, \bibinfo {author} {\bibfnamefont {Y.-K.}\ \bibnamefont {Liu}},
  \bibinfo {author} {\bibfnamefont {S.~T.}\ \bibnamefont {Flammia}}, \bibinfo
  {author} {\bibfnamefont {S.}~\bibnamefont {Becker}}, \ and\ \bibinfo {author}
  {\bibfnamefont {J.}~\bibnamefont {Eisert}},\ }\href {\doibase
  10.1103/PhysRevLett.105.150401} {\bibfield  {journal} {\bibinfo  {journal}
  {Phys. Rev. Lett.}\ }\textbf {\bibinfo {volume} {105}},\ \bibinfo {pages}
  {150401} (\bibinfo {year} {2010})}\BibitemShut {NoStop}%
\bibitem [{\citenamefont {Cramer}\ \emph {et~al.}(2010)\citenamefont {Cramer},
  \citenamefont {Plenio}, \citenamefont {Flammia}, \citenamefont {Somma},
  \citenamefont {Gross}, \citenamefont {Bartlett}, \citenamefont
  {Landon-Cardinal}, \citenamefont {Poulin},\ and\ \citenamefont
  {Liu}}]{Cramer2010}%
  \BibitemOpen
  \bibfield  {author} {\bibinfo {author} {\bibfnamefont {M.}~\bibnamefont
  {Cramer}}, \bibinfo {author} {\bibfnamefont {M.~B.}\ \bibnamefont {Plenio}},
  \bibinfo {author} {\bibfnamefont {S.~T.}\ \bibnamefont {Flammia}}, \bibinfo
  {author} {\bibfnamefont {R.}~\bibnamefont {Somma}}, \bibinfo {author}
  {\bibfnamefont {D.}~\bibnamefont {Gross}}, \bibinfo {author} {\bibfnamefont
  {S.~D.}\ \bibnamefont {Bartlett}}, \bibinfo {author} {\bibfnamefont
  {O.}~\bibnamefont {Landon-Cardinal}}, \bibinfo {author} {\bibfnamefont
  {D.}~\bibnamefont {Poulin}}, \ and\ \bibinfo {author} {\bibfnamefont {Y.-K.}\
  \bibnamefont {Liu}},\ }\href {\doibase 10.1038/ncomms1147} {\bibfield
  {journal} {\bibinfo  {journal} {Nature Communications}\ }\textbf {\bibinfo
  {volume} {1}},\ \bibinfo {pages} {149} (\bibinfo {year} {2010})}\BibitemShut
  {NoStop}%
\bibitem [{\citenamefont {Torlai}\ \emph {et~al.}(2018)\citenamefont {Torlai},
  \citenamefont {Mazzola}, \citenamefont {Carrasquilla}, \citenamefont
  {Troyer}, \citenamefont {Melko},\ and\ \citenamefont {Carleo}}]{Carleo2018}%
  \BibitemOpen
  \bibfield  {author} {\bibinfo {author} {\bibfnamefont {G.}~\bibnamefont
  {Torlai}}, \bibinfo {author} {\bibfnamefont {G.}~\bibnamefont {Mazzola}},
  \bibinfo {author} {\bibfnamefont {J.}~\bibnamefont {Carrasquilla}}, \bibinfo
  {author} {\bibfnamefont {M.}~\bibnamefont {Troyer}}, \bibinfo {author}
  {\bibfnamefont {R.}~\bibnamefont {Melko}}, \ and\ \bibinfo {author}
  {\bibfnamefont {G.}~\bibnamefont {Carleo}},\ }\href {\doibase
  10.1038/s41567-018-0048-5} {\bibfield  {journal} {\bibinfo  {journal} {Nature
  Physics}\ }\textbf {\bibinfo {volume} {14}},\ \bibinfo {pages} {447}
  (\bibinfo {year} {2018})}\BibitemShut {NoStop}%
\bibitem [{\citenamefont {Carrasquilla}\ \emph {et~al.}(2019)\citenamefont
  {Carrasquilla}, \citenamefont {Torlai}, \citenamefont {Melko},\ and\
  \citenamefont {Aolita}}]{Carrasquilla2019}%
  \BibitemOpen
  \bibfield  {author} {\bibinfo {author} {\bibfnamefont {J.}~\bibnamefont
  {Carrasquilla}}, \bibinfo {author} {\bibfnamefont {G.}~\bibnamefont
  {Torlai}}, \bibinfo {author} {\bibfnamefont {R.~G.}\ \bibnamefont {Melko}}, \
  and\ \bibinfo {author} {\bibfnamefont {L.}~\bibnamefont {Aolita}},\ }\href
  {\doibase 10.1038/s42256-019-0028-1} {\bibfield  {journal} {\bibinfo
  {journal} {Nature Machine Intelligence}\ }\textbf {\bibinfo {volume} {1}},\
  \bibinfo {pages} {155} (\bibinfo {year} {2019})}\BibitemShut {NoStop}%
\bibitem [{\citenamefont {Tiunov}\ \emph {et~al.}(2020)\citenamefont {Tiunov},
  \citenamefont {(Vyborova)}, \citenamefont {Ulanov}, \citenamefont {Lvovsky},\
  and\ \citenamefont {Fedorov}}]{Tiunov2020}%
  \BibitemOpen
  \bibfield  {author} {\bibinfo {author} {\bibfnamefont {E.~S.}\ \bibnamefont
  {Tiunov}}, \bibinfo {author} {\bibfnamefont {V.~V.~T.}\ \bibnamefont
  {(Vyborova)}}, \bibinfo {author} {\bibfnamefont {A.~E.}\ \bibnamefont
  {Ulanov}}, \bibinfo {author} {\bibfnamefont {A.~I.}\ \bibnamefont {Lvovsky}},
  \ and\ \bibinfo {author} {\bibfnamefont {A.~K.}\ \bibnamefont {Fedorov}},\
  }\href {\doibase 10.1364/OPTICA.389482} {\bibfield  {journal} {\bibinfo
  {journal} {Optica}\ }\textbf {\bibinfo {volume} {7}},\ \bibinfo {pages} {448}
  (\bibinfo {year} {2020})}\BibitemShut {NoStop}%
\bibitem [{\citenamefont {Koutn\'y}\ \emph {et~al.}(2022)\citenamefont
  {Koutn\'y}, \citenamefont {Motka}, \citenamefont {Hradil}, \citenamefont
  {\ifmmode \check{R}\else \v{R}\fi{}eh\'a\ifmmode~\check{c}\else
  \v{c}\fi{}ek},\ and\ \citenamefont {S\'anchez-Soto}}]{Hradil2022}%
  \BibitemOpen
  \bibfield  {author} {\bibinfo {author} {\bibfnamefont {D.}~\bibnamefont
  {Koutn\'y}}, \bibinfo {author} {\bibfnamefont {L.}~\bibnamefont {Motka}},
  \bibinfo {author} {\bibfnamefont {Z.~c.~v.}\ \bibnamefont {Hradil}}, \bibinfo
  {author} {\bibfnamefont {J.}~\bibnamefont {\ifmmode \check{R}\else
  \v{R}\fi{}eh\'a\ifmmode~\check{c}\else \v{c}\fi{}ek}}, \ and\ \bibinfo
  {author} {\bibfnamefont {L.~L.}\ \bibnamefont {S\'anchez-Soto}},\ }\href
  {\doibase 10.1103/PhysRevA.106.012409} {\bibfield  {journal} {\bibinfo
  {journal} {Phys. Rev. A}\ }\textbf {\bibinfo {volume} {106}},\ \bibinfo
  {pages} {012409} (\bibinfo {year} {2022})}\BibitemShut {NoStop}%
\bibitem [{\citenamefont {Kurmapu}\ \emph {et~al.}(2023)\citenamefont
  {Kurmapu}, \citenamefont {Tiunova}, \citenamefont {Tiunov}, \citenamefont
  {Ringbauer}, \citenamefont {Maier}, \citenamefont {Blatt}, \citenamefont
  {Monz}, \citenamefont {Fedorov},\ and\ \citenamefont
  {Lvovsky}}]{Lvovsky2022}%
  \BibitemOpen
  \bibfield  {author} {\bibinfo {author} {\bibfnamefont {M.~K.}\ \bibnamefont
  {Kurmapu}}, \bibinfo {author} {\bibfnamefont {V.}~\bibnamefont {Tiunova}},
  \bibinfo {author} {\bibfnamefont {E.}~\bibnamefont {Tiunov}}, \bibinfo
  {author} {\bibfnamefont {M.}~\bibnamefont {Ringbauer}}, \bibinfo {author}
  {\bibfnamefont {C.}~\bibnamefont {Maier}}, \bibinfo {author} {\bibfnamefont
  {R.}~\bibnamefont {Blatt}}, \bibinfo {author} {\bibfnamefont
  {T.}~\bibnamefont {Monz}}, \bibinfo {author} {\bibfnamefont {A.~K.}\
  \bibnamefont {Fedorov}}, \ and\ \bibinfo {author} {\bibfnamefont
  {A.}~\bibnamefont {Lvovsky}},\ }\href {\doibase 10.1103/PRXQuantum.4.040345}
  {\bibfield  {journal} {\bibinfo  {journal} {PRX Quantum}\ }\textbf {\bibinfo
  {volume} {4}},\ \bibinfo {pages} {040345} (\bibinfo {year}
  {2023})}\BibitemShut {NoStop}%
\bibitem [{\citenamefont {Bogdanov}(2009)}]{Bogdanov2009}%
  \BibitemOpen
  \bibfield  {author} {\bibinfo {author} {\bibfnamefont {Y.~I.}\ \bibnamefont
  {Bogdanov}},\ }\href {\doibase 10.1134/S106377610906003X} {\bibfield
  {journal} {\bibinfo  {journal} {Journal of Experimental and Theoretical
  Physics}\ }\textbf {\bibinfo {volume} {108}},\ \bibinfo {pages} {928}
  (\bibinfo {year} {2009})}\BibitemShut {NoStop}%
\bibitem [{\citenamefont {Flammia}\ and\ \citenamefont
  {Liu}(2011)}]{Flammia2011}%
  \BibitemOpen
  \bibfield  {author} {\bibinfo {author} {\bibfnamefont {S.~T.}\ \bibnamefont
  {Flammia}}\ and\ \bibinfo {author} {\bibfnamefont {Y.-K.}\ \bibnamefont
  {Liu}},\ }\href {\doibase 10.1103/PhysRevLett.106.230501} {\bibfield
  {journal} {\bibinfo  {journal} {Phys. Rev. Lett.}\ }\textbf {\bibinfo
  {volume} {106}},\ \bibinfo {pages} {230501} (\bibinfo {year}
  {2011})}\BibitemShut {NoStop}%
\bibitem [{\citenamefont {da~Silva}\ \emph {et~al.}(2011)\citenamefont
  {da~Silva}, \citenamefont {Landon-Cardinal},\ and\ \citenamefont
  {Poulin}}]{Silva2011}%
  \BibitemOpen
  \bibfield  {author} {\bibinfo {author} {\bibfnamefont {M.~P.}\ \bibnamefont
  {da~Silva}}, \bibinfo {author} {\bibfnamefont {O.}~\bibnamefont
  {Landon-Cardinal}}, \ and\ \bibinfo {author} {\bibfnamefont {D.}~\bibnamefont
  {Poulin}},\ }\href {\doibase 10.1103/PhysRevLett.107.210404} {\bibfield
  {journal} {\bibinfo  {journal} {Phys. Rev. Lett.}\ }\textbf {\bibinfo
  {volume} {107}},\ \bibinfo {pages} {210404} (\bibinfo {year}
  {2011})}\BibitemShut {NoStop}%
\bibitem [{\citenamefont {Blume-Kohout}(2012)}]{Blume-Kohout2012}%
  \BibitemOpen
  \bibfield  {author} {\bibinfo {author} {\bibfnamefont {R.}~\bibnamefont
  {Blume-Kohout}},\ }\href@noop {} {\enquote {\bibinfo {title} {Robust error
  bars for quantum tomography},}\ } (\bibinfo {year} {2012}),\ \Eprint
  {http://arxiv.org/abs/1202.5270} {arXiv:1202.5270 [quant-ph]} \BibitemShut
  {NoStop}%
\bibitem [{\citenamefont {Christandl}\ and\ \citenamefont
  {Renner}(2012)}]{Renner2012}%
  \BibitemOpen
  \bibfield  {author} {\bibinfo {author} {\bibfnamefont {M.}~\bibnamefont
  {Christandl}}\ and\ \bibinfo {author} {\bibfnamefont {R.}~\bibnamefont
  {Renner}},\ }\href {\doibase 10.1103/PhysRevLett.109.120403} {\bibfield
  {journal} {\bibinfo  {journal} {Phys. Rev. Lett.}\ }\textbf {\bibinfo
  {volume} {109}},\ \bibinfo {pages} {120403} (\bibinfo {year}
  {2012})}\BibitemShut {NoStop}%
\bibitem [{\citenamefont {Bellini}\ \emph {et~al.}(2012)\citenamefont
  {Bellini}, \citenamefont {Coelho}, \citenamefont {Filippov}, \citenamefont
  {Man'ko},\ and\ \citenamefont {Zavatta}}]{Filippov2012}%
  \BibitemOpen
  \bibfield  {author} {\bibinfo {author} {\bibfnamefont {M.}~\bibnamefont
  {Bellini}}, \bibinfo {author} {\bibfnamefont {A.~S.}\ \bibnamefont {Coelho}},
  \bibinfo {author} {\bibfnamefont {S.~N.}\ \bibnamefont {Filippov}}, \bibinfo
  {author} {\bibfnamefont {V.~I.}\ \bibnamefont {Man'ko}}, \ and\ \bibinfo
  {author} {\bibfnamefont {A.}~\bibnamefont {Zavatta}},\ }\href {\doibase
  10.1103/PhysRevA.85.052129} {\bibfield  {journal} {\bibinfo  {journal} {Phys.
  Rev. A}\ }\textbf {\bibinfo {volume} {85}},\ \bibinfo {pages} {052129}
  (\bibinfo {year} {2012})}\BibitemShut {NoStop}%
\bibitem [{\citenamefont {Flammia}\ \emph {et~al.}(2012)\citenamefont
  {Flammia}, \citenamefont {Gross}, \citenamefont {Liu},\ and\ \citenamefont
  {Eisert}}]{Flammia2012}%
  \BibitemOpen
  \bibfield  {author} {\bibinfo {author} {\bibfnamefont {S.~T.}\ \bibnamefont
  {Flammia}}, \bibinfo {author} {\bibfnamefont {D.}~\bibnamefont {Gross}},
  \bibinfo {author} {\bibfnamefont {Y.-K.}\ \bibnamefont {Liu}}, \ and\
  \bibinfo {author} {\bibfnamefont {J.}~\bibnamefont {Eisert}},\ }\href
  {\doibase 10.1088/1367-2630/14/9/095022} {\bibfield  {journal} {\bibinfo
  {journal} {New Journal of Physics}\ }\textbf {\bibinfo {volume} {14}},\
  \bibinfo {pages} {095022} (\bibinfo {year} {2012})}\BibitemShut {NoStop}%
\bibitem [{\citenamefont {Sugiyama}\ \emph {et~al.}(2013)\citenamefont
  {Sugiyama}, \citenamefont {Turner},\ and\ \citenamefont
  {Murao}}]{Sugiyama2013}%
  \BibitemOpen
  \bibfield  {author} {\bibinfo {author} {\bibfnamefont {T.}~\bibnamefont
  {Sugiyama}}, \bibinfo {author} {\bibfnamefont {P.~S.}\ \bibnamefont
  {Turner}}, \ and\ \bibinfo {author} {\bibfnamefont {M.}~\bibnamefont
  {Murao}},\ }\href {\doibase 10.1103/PhysRevLett.111.160406} {\bibfield
  {journal} {\bibinfo  {journal} {Phys. Rev. Lett.}\ }\textbf {\bibinfo
  {volume} {111}},\ \bibinfo {pages} {160406} (\bibinfo {year}
  {2013})}\BibitemShut {NoStop}%
\bibitem [{\citenamefont {Faist}\ and\ \citenamefont
  {Renner}(2016)}]{Renner2016}%
  \BibitemOpen
  \bibfield  {author} {\bibinfo {author} {\bibfnamefont {P.}~\bibnamefont
  {Faist}}\ and\ \bibinfo {author} {\bibfnamefont {R.}~\bibnamefont {Renner}},\
  }\href {\doibase 10.1103/PhysRevLett.117.010404} {\bibfield  {journal}
  {\bibinfo  {journal} {Phys. Rev. Lett.}\ }\textbf {\bibinfo {volume} {117}},\
  \bibinfo {pages} {010404} (\bibinfo {year} {2016})}\BibitemShut {NoStop}%
\bibitem [{\citenamefont {Wang}\ \emph {et~al.}(2019)\citenamefont {Wang},
  \citenamefont {Scholz},\ and\ \citenamefont {Renner}}]{Renner2019}%
  \BibitemOpen
  \bibfield  {author} {\bibinfo {author} {\bibfnamefont {J.}~\bibnamefont
  {Wang}}, \bibinfo {author} {\bibfnamefont {V.~B.}\ \bibnamefont {Scholz}}, \
  and\ \bibinfo {author} {\bibfnamefont {R.}~\bibnamefont {Renner}},\ }\href
  {\doibase 10.1103/PhysRevLett.122.190401} {\bibfield  {journal} {\bibinfo
  {journal} {Phys. Rev. Lett.}\ }\textbf {\bibinfo {volume} {122}},\ \bibinfo
  {pages} {190401} (\bibinfo {year} {2019})}\BibitemShut {NoStop}%
\bibitem [{\citenamefont {Kiktenko}\ \emph {et~al.}(2020)\citenamefont
  {Kiktenko}, \citenamefont {Kublikova},\ and\ \citenamefont
  {Fedorov}}]{Kiktenko20202}%
  \BibitemOpen
  \bibfield  {author} {\bibinfo {author} {\bibfnamefont {E.~O.}\ \bibnamefont
  {Kiktenko}}, \bibinfo {author} {\bibfnamefont {D.~N.}\ \bibnamefont
  {Kublikova}}, \ and\ \bibinfo {author} {\bibfnamefont {A.}~\bibnamefont
  {Fedorov}},\ }\href {\doibase 10.1117/1.oe.59.6.061614} {\bibfield  {journal}
  {\bibinfo  {journal} {Optical Engineering}\ }\textbf {\bibinfo {volume}
  {59}},\ \bibinfo {pages} {061614} (\bibinfo {year} {2020})}\BibitemShut
  {NoStop}%
\bibitem [{\citenamefont {Kiktenko}\ \emph {et~al.}(2021)\citenamefont
  {Kiktenko}, \citenamefont {Norkin},\ and\ \citenamefont
  {Fedorov}}]{Kiktenko2021}%
  \BibitemOpen
  \bibfield  {author} {\bibinfo {author} {\bibfnamefont {E.~O.}\ \bibnamefont
  {Kiktenko}}, \bibinfo {author} {\bibfnamefont {D.~O.}\ \bibnamefont
  {Norkin}}, \ and\ \bibinfo {author} {\bibfnamefont {A.~K.}\ \bibnamefont
  {Fedorov}},\ }\href {\doibase 10.1088/1367-2630/ac3cf7} {\bibfield  {journal}
  {\bibinfo  {journal} {New Journal of Physics}\ }\textbf {\bibinfo {volume}
  {23}},\ \bibinfo {pages} {123022} (\bibinfo {year} {2021})}\BibitemShut
  {NoStop}%
\bibitem [{\citenamefont {Poyatos}\ \emph {et~al.}(1997)\citenamefont
  {Poyatos}, \citenamefont {Cirac},\ and\ \citenamefont
  {Zoller}}]{Poyatos1997}%
  \BibitemOpen
  \bibfield  {author} {\bibinfo {author} {\bibfnamefont {J.~F.}\ \bibnamefont
  {Poyatos}}, \bibinfo {author} {\bibfnamefont {J.~I.}\ \bibnamefont {Cirac}},
  \ and\ \bibinfo {author} {\bibfnamefont {P.}~\bibnamefont {Zoller}},\ }\href
  {\doibase 10.1103/PhysRevLett.78.390} {\bibfield  {journal} {\bibinfo
  {journal} {Phys. Rev. Lett.}\ }\textbf {\bibinfo {volume} {78}},\ \bibinfo
  {pages} {390} (\bibinfo {year} {1997})}\BibitemShut {NoStop}%
\bibitem [{\citenamefont {Chuang}\ and\ \citenamefont
  {Nielsen}(1997)}]{Chuang1997}%
  \BibitemOpen
  \bibfield  {author} {\bibinfo {author} {\bibfnamefont {I.~L.}\ \bibnamefont
  {Chuang}}\ and\ \bibinfo {author} {\bibfnamefont {M.~A.}\ \bibnamefont
  {Nielsen}},\ }\href {\doibase 10.1080/09500349708231894} {\bibfield
  {journal} {\bibinfo  {journal} {Journal of Modern Optics}\ }\textbf {\bibinfo
  {volume} {44}},\ \bibinfo {pages} {2455} (\bibinfo {year}
  {1997})}\BibitemShut {NoStop}%
\bibitem [{\citenamefont {D'Ariano}\ and\ \citenamefont
  {Lo~Presti}(2001)}]{DAriano2001}%
  \BibitemOpen
  \bibfield  {author} {\bibinfo {author} {\bibfnamefont {G.~M.}\ \bibnamefont
  {D'Ariano}}\ and\ \bibinfo {author} {\bibfnamefont {P.}~\bibnamefont
  {Lo~Presti}},\ }\href {\doibase 10.1103/PhysRevLett.86.4195} {\bibfield
  {journal} {\bibinfo  {journal} {Phys. Rev. Lett.}\ }\textbf {\bibinfo
  {volume} {86}},\ \bibinfo {pages} {4195} (\bibinfo {year}
  {2001})}\BibitemShut {NoStop}%
\bibitem [{\citenamefont {O'Brien}\ \emph {et~al.}(2004)\citenamefont
  {O'Brien}, \citenamefont {Pryde}, \citenamefont {Gilchrist}, \citenamefont
  {James}, \citenamefont {Langford}, \citenamefont {Ralph},\ and\ \citenamefont
  {White}}]{White2004}%
  \BibitemOpen
  \bibfield  {author} {\bibinfo {author} {\bibfnamefont {J.~L.}\ \bibnamefont
  {O'Brien}}, \bibinfo {author} {\bibfnamefont {G.~J.}\ \bibnamefont {Pryde}},
  \bibinfo {author} {\bibfnamefont {A.}~\bibnamefont {Gilchrist}}, \bibinfo
  {author} {\bibfnamefont {D.~F.~V.}\ \bibnamefont {James}}, \bibinfo {author}
  {\bibfnamefont {N.~K.}\ \bibnamefont {Langford}}, \bibinfo {author}
  {\bibfnamefont {T.~C.}\ \bibnamefont {Ralph}}, \ and\ \bibinfo {author}
  {\bibfnamefont {A.~G.}\ \bibnamefont {White}},\ }\href {\doibase
  10.1103/PhysRevLett.93.080502} {\bibfield  {journal} {\bibinfo  {journal}
  {Phys. Rev. Lett.}\ }\textbf {\bibinfo {volume} {93}},\ \bibinfo {pages}
  {080502} (\bibinfo {year} {2004})}\BibitemShut {NoStop}%
\bibitem [{\citenamefont {Knill}\ \emph {et~al.}(2008)\citenamefont {Knill},
  \citenamefont {Leibfried}, \citenamefont {Reichle}, \citenamefont {Britton},
  \citenamefont {Blakestad}, \citenamefont {Jost}, \citenamefont {Langer},
  \citenamefont {Ozeri}, \citenamefont {Seidelin},\ and\ \citenamefont
  {Wineland}}]{Wineland2008}%
  \BibitemOpen
  \bibfield  {author} {\bibinfo {author} {\bibfnamefont {E.}~\bibnamefont
  {Knill}}, \bibinfo {author} {\bibfnamefont {D.}~\bibnamefont {Leibfried}},
  \bibinfo {author} {\bibfnamefont {R.}~\bibnamefont {Reichle}}, \bibinfo
  {author} {\bibfnamefont {J.}~\bibnamefont {Britton}}, \bibinfo {author}
  {\bibfnamefont {R.~B.}\ \bibnamefont {Blakestad}}, \bibinfo {author}
  {\bibfnamefont {J.~D.}\ \bibnamefont {Jost}}, \bibinfo {author}
  {\bibfnamefont {C.}~\bibnamefont {Langer}}, \bibinfo {author} {\bibfnamefont
  {R.}~\bibnamefont {Ozeri}}, \bibinfo {author} {\bibfnamefont
  {S.}~\bibnamefont {Seidelin}}, \ and\ \bibinfo {author} {\bibfnamefont
  {D.~J.}\ \bibnamefont {Wineland}},\ }\href {\doibase
  10.1103/PhysRevA.77.012307} {\bibfield  {journal} {\bibinfo  {journal} {Phys.
  Rev. A}\ }\textbf {\bibinfo {volume} {77}},\ \bibinfo {pages} {012307}
  (\bibinfo {year} {2008})}\BibitemShut {NoStop}%
\bibitem [{\citenamefont {Watson}(1967)}]{Watson1967}%
  \BibitemOpen
  \bibfield  {author} {\bibinfo {author} {\bibfnamefont {G.~S.}\ \bibnamefont
  {Watson}},\ }\href {\doibase 10.1214/aoms/1177698603} {\bibfield  {journal}
  {\bibinfo  {journal} {The Annals of Mathematical Statistics}\ }\textbf
  {\bibinfo {volume} {38}},\ \bibinfo {pages} {1679 } (\bibinfo {year}
  {1967})}\BibitemShut {NoStop}%
\bibitem [{\citenamefont {Kjeldsen}(1993)}]{Kjeldsen1993}%
  \BibitemOpen
  \bibfield  {author} {\bibinfo {author} {\bibfnamefont {T.~H.}\ \bibnamefont
  {Kjeldsen}},\ }\href {\doibase https://doi.org/10.1006/hmat.1993.1004}
  {\bibfield  {journal} {\bibinfo  {journal} {Historia Mathematica}\ }\textbf
  {\bibinfo {volume} {20}},\ \bibinfo {pages} {19} (\bibinfo {year}
  {1993})}\BibitemShut {NoStop}%
\bibitem [{\citenamefont {Lin}(2017)}]{Lin2017}%
  \BibitemOpen
  \bibfield  {author} {\bibinfo {author} {\bibfnamefont {G.~D.}\ \bibnamefont
  {Lin}},\ }\href {\doibase 10.1186/s40488-017-0059-2} {\bibfield  {journal}
  {\bibinfo  {journal} {Journal of Statistical Distributions and Applications}\
  }\textbf {\bibinfo {volume} {4}},\ \bibinfo {pages} {5} (\bibinfo {year}
  {2017})}\BibitemShut {NoStop}%
\bibitem [{\citenamefont {Newcomer}\ \emph {et~al.}(2008)\citenamefont
  {Newcomer}, \citenamefont {Neerchal},\ and\ \citenamefont
  {Morel}}]{newcomer2008computation}%
  \BibitemOpen
  \bibfield  {author} {\bibinfo {author} {\bibfnamefont {J.}~\bibnamefont
  {Newcomer}}, \bibinfo {author} {\bibfnamefont {N.}~\bibnamefont {Neerchal}},
  \ and\ \bibinfo {author} {\bibfnamefont {J.}~\bibnamefont {Morel}},\
  }\href@noop {} {\bibfield  {journal} {\bibinfo  {journal} {Department of
  Mathematics and Statistics, University of Maryland, Baltimore, USA}\ }
  (\bibinfo {year} {2008})}\BibitemShut {NoStop}%
\bibitem [{\citenamefont {Ouimet}(2021)}]{stats4010002}%
  \BibitemOpen
  \bibfield  {author} {\bibinfo {author} {\bibfnamefont {F.}~\bibnamefont
  {Ouimet}},\ }\href {\doibase 10.3390/stats4010002} {\bibfield  {journal}
  {\bibinfo  {journal} {Stats}\ }\textbf {\bibinfo {volume} {4}},\ \bibinfo
  {pages} {18} (\bibinfo {year} {2021})}\BibitemShut {NoStop}%
\bibitem [{\citenamefont {Bodenham}\ and\ \citenamefont
  {Adams}(2016)}]{bodenham2016comparison}%
  \BibitemOpen
  \bibfield  {author} {\bibinfo {author} {\bibfnamefont {D.~A.}\ \bibnamefont
  {Bodenham}}\ and\ \bibinfo {author} {\bibfnamefont {N.~M.}\ \bibnamefont
  {Adams}},\ }\href@noop {} {\bibfield  {journal} {\bibinfo  {journal}
  {Statistics and Computing}\ }\textbf {\bibinfo {volume} {26}},\ \bibinfo
  {pages} {917} (\bibinfo {year} {2016})}\BibitemShut {NoStop}%
\bibitem [{\citenamefont {Efron}\ and\ \citenamefont
  {Tibshirani}(1993)}]{EfroTibs93}%
  \BibitemOpen
  \bibfield  {author} {\bibinfo {author} {\bibfnamefont {B.}~\bibnamefont
  {Efron}}\ and\ \bibinfo {author} {\bibfnamefont {R.~J.}\ \bibnamefont
  {Tibshirani}},\ }\href@noop {} {\emph {\bibinfo {title} {An Introduction to
  the Bootstrap}}},\ \bibinfo {series} {Monographs on Statistics and Applied
  Probability}\ No.~\bibinfo {number} {57}\ (\bibinfo  {publisher} {Chapman \&
  Hall/CRC},\ \bibinfo {address} {Boca Raton, Florida, USA},\ \bibinfo {year}
  {1993})\BibitemShut {NoStop}%
\bibitem [{\citenamefont {Jamio{\l}kowski}(1972)}]{Jamiokowski1972}%
  \BibitemOpen
  \bibfield  {author} {\bibinfo {author} {\bibfnamefont {A.}~\bibnamefont
  {Jamio{\l}kowski}},\ }\href {\doibase
  https://doi.org/10.1016/0034-4877(72)90011-0} {\bibfield  {journal} {\bibinfo
   {journal} {Reports on Mathematical Physics}\ }\textbf {\bibinfo {volume}
  {3}},\ \bibinfo {pages} {275} (\bibinfo {year} {1972})}\BibitemShut {NoStop}%
\bibitem [{\citenamefont {Choi}(1975)}]{Choi1975}%
  \BibitemOpen
  \bibfield  {author} {\bibinfo {author} {\bibfnamefont {M.-D.}\ \bibnamefont
  {Choi}},\ }\href {\doibase https://doi.org/10.1016/0024-3795(75)90075-0}
  {\bibfield  {journal} {\bibinfo  {journal} {Linear Algebra and its
  Applications}\ }\textbf {\bibinfo {volume} {10}},\ \bibinfo {pages} {285}
  (\bibinfo {year} {1975})}\BibitemShut {NoStop}%
\bibitem [{\citenamefont {Jiang}\ \emph {et~al.}(2013)\citenamefont {Jiang},
  \citenamefont {Luo},\ and\ \citenamefont {Fu}}]{Jiang2013}%
  \BibitemOpen
  \bibfield  {author} {\bibinfo {author} {\bibfnamefont {M.}~\bibnamefont
  {Jiang}}, \bibinfo {author} {\bibfnamefont {S.}~\bibnamefont {Luo}}, \ and\
  \bibinfo {author} {\bibfnamefont {S.}~\bibnamefont {Fu}},\ }\href {\doibase
  10.1103/PhysRevA.87.022310} {\bibfield  {journal} {\bibinfo  {journal} {Phys.
  Rev. A}\ }\textbf {\bibinfo {volume} {87}},\ \bibinfo {pages} {022310}
  (\bibinfo {year} {2013})}\BibitemShut {NoStop}%
\bibitem [{\citenamefont {Zyczkowski}\ and\ \citenamefont
  {Sommers}(2001)}]{zyczkowski2001induced}%
  \BibitemOpen
  \bibfield  {author} {\bibinfo {author} {\bibfnamefont {K.}~\bibnamefont
  {Zyczkowski}}\ and\ \bibinfo {author} {\bibfnamefont {H.-J.}\ \bibnamefont
  {Sommers}},\ }\href
  {https://iopscience.iop.org/article/10.1088/0305-4470/34/35/335} {\bibfield
  {journal} {\bibinfo  {journal} {Journal of Physics A: Mathematical and
  General}\ }\textbf {\bibinfo {volume} {34}},\ \bibinfo {pages} {7111}
  (\bibinfo {year} {2001})}\BibitemShut {NoStop}%
\bibitem [{IBM()}]{IBMQ5}%
  \BibitemOpen
  \href@noop {} {\enquote {\bibinfo {title} {Ibm q 5 yorktown v1.x.x version
  log},}\ }\bibinfo {howpublished} {\url{
  https://github.com/adcorcol/ibmqx-backend-information/blob/master/backends/yorktown/V1/README.md}}\BibitemShut
  {NoStop}%
\bibitem [{\citenamefont {Lobino}\ \emph {et~al.}(2008)\citenamefont {Lobino},
  \citenamefont {Korystov}, \citenamefont {Kupchak}, \citenamefont {Figueroa},
  \citenamefont {Sanders},\ and\ \citenamefont {Lvovsky}}]{Lvovsky2008}%
  \BibitemOpen
  \bibfield  {author} {\bibinfo {author} {\bibfnamefont {M.}~\bibnamefont
  {Lobino}}, \bibinfo {author} {\bibfnamefont {D.}~\bibnamefont {Korystov}},
  \bibinfo {author} {\bibfnamefont {C.}~\bibnamefont {Kupchak}}, \bibinfo
  {author} {\bibfnamefont {E.}~\bibnamefont {Figueroa}}, \bibinfo {author}
  {\bibfnamefont {B.~C.}\ \bibnamefont {Sanders}}, \ and\ \bibinfo {author}
  {\bibfnamefont {A.~I.}\ \bibnamefont {Lvovsky}},\ }\href {\doibase
  10.1126/science.1162086} {\bibfield  {journal} {\bibinfo  {journal}
  {Science}\ }\textbf {\bibinfo {volume} {322}},\ \bibinfo {pages} {563}
  (\bibinfo {year} {2008})}\BibitemShut {NoStop}%
\bibitem [{\citenamefont {Rahimi-Keshari}\ \emph {et~al.}(2011)\citenamefont
  {Rahimi-Keshari}, \citenamefont {Scherer}, \citenamefont {Mann},
  \citenamefont {Rezakhani}, \citenamefont {Lvovsky},\ and\ \citenamefont
  {Sanders}}]{Lvovsky2011}%
  \BibitemOpen
  \bibfield  {author} {\bibinfo {author} {\bibfnamefont {S.}~\bibnamefont
  {Rahimi-Keshari}}, \bibinfo {author} {\bibfnamefont {A.}~\bibnamefont
  {Scherer}}, \bibinfo {author} {\bibfnamefont {A.}~\bibnamefont {Mann}},
  \bibinfo {author} {\bibfnamefont {A.~T.}\ \bibnamefont {Rezakhani}}, \bibinfo
  {author} {\bibfnamefont {A.~I.}\ \bibnamefont {Lvovsky}}, \ and\ \bibinfo
  {author} {\bibfnamefont {B.~C.}\ \bibnamefont {Sanders}},\ }\href {\doibase
  10.1088/1367-2630/13/1/013006} {\bibfield  {journal} {\bibinfo  {journal}
  {New Journal of Physics}\ }\textbf {\bibinfo {volume} {13}},\ \bibinfo
  {pages} {013006} (\bibinfo {year} {2011})}\BibitemShut {NoStop}%
\bibitem [{\citenamefont {Anis}\ and\ \citenamefont
  {Lvovsky}(2012)}]{Anis2012}%
  \BibitemOpen
  \bibfield  {author} {\bibinfo {author} {\bibfnamefont {A.}~\bibnamefont
  {Anis}}\ and\ \bibinfo {author} {\bibfnamefont {A.~I.}\ \bibnamefont
  {Lvovsky}},\ }\href {\doibase 10.1088/1367-2630/14/10/105021} {\bibfield
  {journal} {\bibinfo  {journal} {New Journal of Physics}\ }\textbf {\bibinfo
  {volume} {14}},\ \bibinfo {pages} {105021} (\bibinfo {year}
  {2012})}\BibitemShut {NoStop}%
\bibitem [{\citenamefont {Fedorov}\ \emph {et~al.}(2015)\citenamefont
  {Fedorov}, \citenamefont {Fedorov}, \citenamefont {Kurochkin},\ and\
  \citenamefont {Lvovsky}}]{Fedorov2015}%
  \BibitemOpen
  \bibfield  {author} {\bibinfo {author} {\bibfnamefont {I.~A.}\ \bibnamefont
  {Fedorov}}, \bibinfo {author} {\bibfnamefont {A.~K.}\ \bibnamefont
  {Fedorov}}, \bibinfo {author} {\bibfnamefont {Y.~V.}\ \bibnamefont
  {Kurochkin}}, \ and\ \bibinfo {author} {\bibfnamefont {A.~I.}\ \bibnamefont
  {Lvovsky}},\ }\href {\doibase 10.1088/1367-2630/17/4/043063} {\bibfield
  {journal} {\bibinfo  {journal} {New Journal of Physics}\ }\textbf {\bibinfo
  {volume} {17}},\ \bibinfo {pages} {043063} (\bibinfo {year}
  {2015})}\BibitemShut {NoStop}%
\bibitem [{\citenamefont {Ghalaii}\ and\ \citenamefont
  {Rezakhani}(2017)}]{ghalaii2017scheme}%
  \BibitemOpen
  \bibfield  {author} {\bibinfo {author} {\bibfnamefont {M.}~\bibnamefont
  {Ghalaii}}\ and\ \bibinfo {author} {\bibfnamefont {A.~T.}\ \bibnamefont
  {Rezakhani}},\ }\href
  {https://journals.aps.org/pra/abstract/10.1103/PhysRevA.95.032336} {\bibfield
   {journal} {\bibinfo  {journal} {Phys. Rev. A}\ }\textbf {\bibinfo {volume}
  {95}},\ \bibinfo {pages} {032336} (\bibinfo {year} {2017})}\BibitemShut
  {NoStop}%
\bibitem [{\citenamefont {Kiktenko}(2023)}]{kiktenko2023exploring}%
  \BibitemOpen
  \bibfield  {author} {\bibinfo {author} {\bibfnamefont {E.~O.}\ \bibnamefont
  {Kiktenko}},\ }\href
  {https://journals.aps.org/pra/abstract/10.1103/PhysRevA.107.032419}
  {\bibfield  {journal} {\bibinfo  {journal} {Phys. Rev. A}\ }\textbf {\bibinfo
  {volume} {107}},\ \bibinfo {pages} {032419} (\bibinfo {year}
  {2023})}\BibitemShut {NoStop}%
\end{thebibliography}%

\end{document}